\documentclass[12pt]{amsart}
\usepackage{geometry}                
\usepackage{graphicx}
\usepackage{amssymb}
\usepackage{mathtools}
\usepackage{epstopdf}

\usepackage{color}
\usepackage[round]{natbib}
\newcommand{\DS}{\emph{ DeepSeasons}\,}

\title{\DS: a Deep Learning scale-selecting approach to Seasonal Forecasts} 
 \author{A. Navarra}
\address{Centro Euromediterraneo sui Cambiamenti Climatici, Bologna, Italy and Universita' di Bologna, Dipartimento di Scienze Geologiche, Biologiche e Ambientali, Bologna, Italy}
\email{antonio.navarra@unibo.it}
\thanks{Corresponding author.} 

\author{G.G. Navarra}
\address{Princeton University, Princeton, NJ, US}

\begin{document}

\begin{abstract}
Seasonal forecasting remains challenging due to the inherent chaotic nature of atmospheric dynamics. This paper introduces \DS, a novel deep learning approach designed to enhance the accuracy and reliability of seasonal forecasts. Leveraging advanced neural network architectures and extensive historical climatic datasets, \DS identifies complex, nonlinear patterns and dependencies in climate variables with similar or improved skill respcet  GCM-based  forecasting methods, at a significant lower cost. The framework also allow tailored application to specific regions or variables, rather than the overall problem of predicting the entire atmosphere/ocean system.  The proposed methods also allow for direct predictions of anomalies and time-means, opening a new approach to long-term forecasting and highlighting its potential for operational deployment in climate-sensitive sectors. This innovative methodology promises substantial improvements in managing climate-related risks and decision-making processes.
\end{abstract}

\maketitle

\section{Introduction}

Weather and climate forecasts have traditionally relied on numerical weather prediction (NWP) models that solve the governing differential equations of the atmosphere using time-marching schemes.  These models simulate the atmospheric state by advancing it step-by-step in time, capturing the evolution of weather systems with remarkable detail over short to medium ranges. However, for long-range forecasts extending beyond 15 days, the inherent chaotic nature of the atmosphere imposes fundamental limits on predictability. \citep{Lorenz1963, Lorenz_2006,VANKEKEM201838} Consequently, forecast skill at these extended ranges is confined to large-scale temporal and spatial averages rather than specific weather events \citep{Molteni1996}. The dependence of  predictability on spatial and temporal scales has been known for some time.  \citet{Shukla1981} pointed out that monthly means may show higher predictability than instantaneous predictions, and the notion has been empirically exploited in operational set-ups (see, for instance, the Copernicus Climate Change Service, C3S\footnote{ http://copernicus.climate.eu}, or the North American Multi-Model Ensemble\footnote{https://www.ncei.noaa.gov/products/weather-climate-models/north-american-multi-model}. The objectives of these long-range forecasting systems at seasonal scale and beyond was aiming at predicting  monthly and seasonal means, using also hybrid statistical-dynamical methods\citep{Becker2022, ZHOU2024102361, Kirtman2014}.

Predicting such climatological means using traditional differential models poses significant challenges. Over extended periods, small errors in initial conditions or model formulations can amplify exponentially, leading to substantial deviations from observed climate statistics \citep{Slingo2011}. Furthermore, the computational resources required to run high-resolution models for seasonal or annual forecasts are substantial \citep{Palmer2019}. The complexity of atmospheric processes and their interactions with other components of the Earth system further complicate long-term predictions using conventional models \citep{Bauer2015}.

Furthermore, seasonal forecasting models show significant drift due to adjustment processes in the initial condition and deficiencies in the model formulation, often amplified by insufficient spatial and temporal resolution. The response has been to perform large ensemble of hindcasts to define a model climatology that is then subtracted from the forecasts so to compensate the systematic drift.  Validation is then performed between the forecasted anomalies (deviation of time means from the model climatology obtained from the hindcasts) and the observed anomaly (deviation of time means from the observed climatology). This procedure effectively turn the system into an anomaly forecast models.

 It was recognized early that the predicting the total field was more difficult than predicting anomalies and so there were attempts to design a model carrying the anomaly as the variables \citep{Navarra1988}. However, these efforts were unsuccessful since the nonlinearity would prevent a simple approach based on a Reynold-like separation between  the basic state climatology and the anomaly deviation.  Recently, there have been other efforts that have revived the usage of anomalies in verification and forecasting showing the advantages of looking at anomalies rather than the full field \citep{Qian2021}, but no model exist that is exploiting the basic property of anomalies to design a forecasting model designed to target on the anomaly itself.

Machine learning (ML) methods offer a promising alternative for predicting space and time means on monthly to decadal timescales. By leveraging large datasets of historical observations and simulations, ML algorithms can capture complex, nonlinear relationships within the climate system without explicitly solving the underlying differential equations \citep{Reichstein2019}. 

Advances in deep learning, particularly in architectures like recurrent neural networks (RNNs) \citep{Hochreiter1997} and transformers,\citep{Vaswani2017} have shown potential in improving climate predictions by directly modeling the dynamics of the atmosphere.  \citet{Bi:2023aa} introduced  "Pangu-Weather", a 3D high-resolution model that leverages deep learning to provide fast and accurate global weather forecasts at synoptic time scales. Their approach demonstrates substantial improvements in efficiency while maintaining forecast accuracy comparable to traditional NWP models. Similarly,  \citet{Chattopadhyay2020,Keisler2022} explored the use of graph neural networks for global weather forecasting, showing that GNNs can effectively capture the spatial dependencies of atmospheric variables on the Earth's surface. \citep{Pathak2022} proposed \textit{FourCastNet}, utilizing adaptive Fourier Neural Operators for high-resolution forecasting, offering a novel approach to modeling complex atmospheric dynamics with impressive accuracy and computational speed. \citep{Lam2023} has also produced a ML model with superior performance at short and medium range weather forecast. Foundational models are also under development \citep{Bodnar:2025aa} in the effort to generate foundational systems that can serve as the basis for development of downstream applications.

The extension of these techniques to longer time scales or even multiyear climate scale is still unclear. Several studies have explored the application of ML techniques to climate with some encouraging result.  For instance, \citet{Weyn2019}  developed a convolutional neural network (CNN) model that emulates atmospheric dynamics and provides competitive forecasts compared to traditional models.  \citet{Arcomano2020} demonstrated that deep learning models could predict precipitation patterns with reasonable accuracy over extended periods. Moreover, research has shown that ML models can enhance subseasonal to seasonal forecasts by capturing complex teleconnections and climate indices like the El Ni\~no Southern Oscillation (ENSO) \citet{Ham2019, Chattopadhyay2020}. Extension to longer time scales suitable for climate scenarios is a currently active topic of research \citep{Eyring:2024ab, Eyring:2024aa} with several options explored to introduce AI/ML methods also to this problems.

These recent studies have demonstrated the growing role of machine learning in advancing weather predictions. By capturing complex patterns and relationships within extensive datasets, ML models offer promising avenues for improving forecast accuracy on various timescales, from nowcasting to seasonal and even decadal forecasts. The extent to which these results can be replicated on longer scales, such as climate scales, is an active area of research.

In this study, we explore the application of machine learning techniques to directly predict atmospheric patterns beyond the synoptic scale up to 12 months. We exploit the particular property of ML methods that allows us to train a network directly on time-averaged quantities, where we expect to have a different predictability from instantaneous weather pattern. We employ transformer-based architectures, inspired by their success in natural language processing, to model the temporal dependencies in  data \citep{Vaswani2017}. By training these models on extensive datasets, we aim to forecast monthly and seasonal means of key atmospheric variables. We assess the performance of these ML models against traditional  models and evaluate their potential to enhance predictability at extended ranges.

The integration of machine learning (ML) into modeling workflows marks a fundamental shift in paradigm, offering a departure from traditional modeling approaches. Classical numerical models  require detailed and accurate description and treatment of system dynamics, significant computational resources, and rigid frameworks for incorporating new information. In contrast, ML enables flexible, data-driven strategies that adapt to complex, nonlinear relationships often difficult to express analytically. This paradigm broadens the horizon of modeling options, allowing modelers to explore a wider range of inputs, learning architectures, and objectives. In practice, ML facilitates the incorporation of diverse datasets and observational sources without requiring a full mechanistic description of the underlying processes. As a result, modelers can experiment with more focused, purpose-driven approaches that are often faster to develop and update.

\DS  exemplifies how targeted learning architectures can be tailored to specific applications. \DS  offers the flexibility to specialize in particular geographic areas or variables of interest without necessitating a comprehensive representation of all physical variables in the climate system. This selective modeling capability is especially powerful, as it permits  for refinement of the models for high-impact variables while managing computational complexity. Rather than simulate the entire atmospheric state, \DS can focus on learning mappings from several kinds of inputs to specific outputs, capturing relevant signals that influence seasonal variability in the region and variable of interest.

Sea surface temperature (SST) in tropical regions plays a crucial role in global climate dynamics, influencing phenomena such as the ElNino  Southern Oscillation (ENSO), monsoon systems, and tropical cyclone formation. Accurate prediction of tropical SST is essential for climate modeling, weather forecasting, and understanding the broader implications of climate change.  The application of ML and AI techniques to SST prediction has gained significant traction in recent years, driven by the increasing availability of high-resolution satellite data, advances in computational power, and the development of sophisticated deep learning architectures  \citep{Taylor_2022, wang2023convolutionalgrunetworkseasonal,10068549,9504603} These methods have demonstrated the potential to capture non-linear relationships and complex spatiotemporal patterns in SST variations that are as challenging for traditional models to represent  as in the case for weather \citep{Lam2023}.

We present here a new multivariate prediction system for atmospheric variables based on a simple transformer architecture that  nevertheless achieves significant results in terms of performance and it is opening the way to more detailed and advanced model that are being developed. The structure of the paper is organized as follows. Section 2  will describe the data sources utilized in the study, Section 3  will describe \DS. Section 4 will show the result for prediction of global SST, Section 5 will contain results for Europe and North America 2m Temperature and Section 6 will contain some preliminary results for three-month averages forecast of 2m temperature over Europe.  The Conclusion in Section 7 and some ideas for  future work will conclude the paper.

\section{Data}

We are utilizing anomaly monthly means data from the ERA5 reanalysis for both training and verification during the period 1940-2022. 

The ERA5 dataset is a widely-used atmospheric reanalysis dataset produced by the European Centre for Medium-Range Weather Forecasts (ECMWF). It provides comprehensive climate and weather data at a high spatial resolution of 0.25 Degrees on a global scale. The dataset covers the entire globe, offering detailed atmospheric, land, and oceanic variables \citep{ERA5}, available from https://www.ecmwf.int/en/forecasts/datasets/reanalysis-datasets/era5.


\section{\DS}

Recently, a number of groups have reported neural network designs that are capable of performing in short and medium range weather forecasts based on spatial and temporal connections using attention mechanisms both in space and time. In this paper we are using a different approach as the network is designed to predict the SST PCs effectively framing the problem as a multivariate time series forecast problem. 

Deep learning methods for time series forecasting is an active area of research in machine learning \citet{KARTAL2023105675,Chen2023}, but later transformers have also been used for time series forecasting \citet{Liu2024,Wen20236778}, also developing specific loss functions \citet{Jadon2024117, jadon2022}. \DS network is using transformers and is capable of using different input fields, using both full attention  and optionally using a probabilistic attention mechanism as the Informer  \citet{ZHOU2023103886}. 

\DS uses an approach based on transformers, in the flavor developed  by  \citep{ZHOU2023103886}, as the Informer that allows for greater efficiency in the attention mechanism. The Informer is very suited for prediction of time series as it allows for long input sequences and it requires less memory than a standard attention mechanism, however in our application the computation demands are very modest and this advantage is not immediately apparent and therefore we can use the Informer as a standard transformer without using the probability and distilling mechanism. The code developed here is a modification  of the code available from Huggingface\footnote{https://huggingface.co/docs/transformers/main/model\_doc/informer}, based on the original Zhou's code\footnote{Available at https://github.com/zhouhaoyi/Informer2020} and it is available on request from the authors.

\DS neural architecture presented centers on a design specifically tailored for predictive tasks. In essence, this architecture leverages the advantages of self-attention mechanisms to effectively capture both short- and long-range dependencies in sequential data. By introducing a dedicated value embedding and sinusoidal positional encoding, the model ensures that each time step is furnished with both content and positional information. Furthermore, temporal features are added to the input features to encode the information about year, month and season. This approach is particularly advantageous for tasks involving temporal sequences, where identifying both proximate and distant correlations can substantially improve forecasts. The input sequence consist of time series of EOF coefficients of the selected input fields, but different lags can be concatenated resulting in a larger dimension for the feature space. For instance, in the case of having a single input field that retains 15 EOF at each time level, using lags 1 and 2 will result in a rearranging of the data in sequence of 45 EOF at each time level. The final number of features per time level is also increased because fixed time features are added encoding year, month and season information. In the example, the final dimension will then be 48. 
  
At the core of the model resides an Encoder/Decoder structure, which preprocesses input sequences through a direct connection layer, transforming the input sequence of features into the higher dimensional latent representation. Sinusoidal positional embeddings then inject temporal context into these embeddings. Within each encoder layer, an Attention module computes queries, keys, and values in a multi-head arrangement, allowing for selective weighting of important temporal signals. Residual connections and layer normalization help stabilize training and preserve essential features, while a pair of feedforward layers, coupled with activation functions such as ReLU, refine learned representations by introducing nonlinearity.

The Decoder mirrors many of these design principles, employing a parallel value projection and positional embedding strategy for the target sequence. Its self-attention module computes intra-decoder representations, while a separate attention block aligns the decoder evolving representations with those previously extracted by the encoder. Similar to the encoder, each decoder layer integrates residual connections, feedforward transformations, and layer normalization to maintain representational stability. This dual-attention mechanism, self-attention and cross-attention equips the decoder to attend not only to the target sequence context but also to crucial information encoded by the encoder.

Finally, a projection layer will attempt to estimate the parameters of the probability distribution of the forecasts conditioned on the input sequence, therefore generating an ensemble of predictions. Here, a series of linear transformations reshape the  hidden states into the required output dimension. By design, this projection step ensures that the networks learned representations can map to a variety of predictive tasks with minimal overhead. Though various probability distributions can be used, the Gaussian distribution is used throughout all cases in the remainder of the paper. In this first paper we are concentrating on the ensemble mean of the forecasts.

\subsection{Loss function}
The loss function used is the negative log-likelihood as the network is learning the parameters of the probability distribution conditioned on the input sequence. The inference is then obtained sampling from the distribution step by step through greedy inference. For each forecast initial date an ensemble can be produced from the conditional probability distribution and the input sequence.

\begin{figure*}
 \centering
 \includegraphics[width=\textwidth]{ 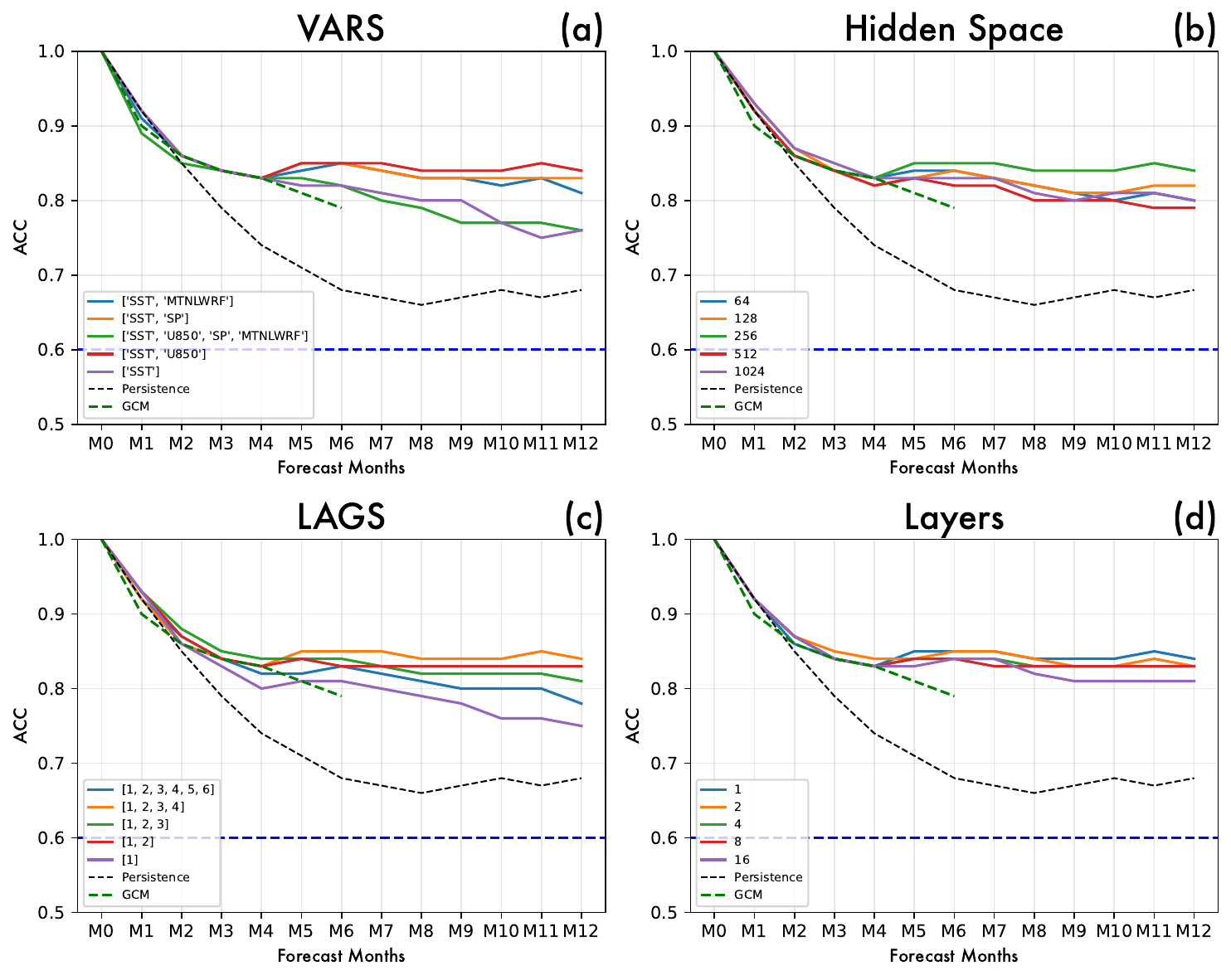}
 \caption{Sensitivity analysis of \DS. Each panel compares the mean forecast skill (spatial anomaly correlation, 0.5-1.0) of the \DS based model versus Persistence and a CMCC GCM over lead times from Month 0 (M0) to Month 12 (M12) for the set of forecasts in the Test set. As a reference the correlation level of 0.6 is indicated by the dashed blue line. The top-left panel (a) examines the input variable combinations (SST, SP, U850, and MTNLWRF) with the best results obtained using either SST and U850 or SST and SP. Using SST and U850, the top-right panel (b) explores the impact of the hidden space dimension (64, 128, 256, 512, and 1024). The bottom left panel (c) investigates the sensitivity to the number of temporal lags, showing that while shorter lags boost short term performance, a tradeoff is achieved with four lags (1,2,3,4). Finally, the bottom-right panel (d) evaluates the influence of network depth (1, 2, 4, 8, and 16 hidden layers), with the final chosen configuration being SST and U850, with 256 hidden space dimension, two transformer layers and four lags at 1,2,3,4 months.} 
\label{Fig2}
\end{figure*}

We have modified the loss function to give some degree of control over the focus of the forecast. Let
\[
\mathbf{x}_t = \left( x_{t,1}, x_{t,2}, \dots, x_{t,N} \right) \in \mathbb{R}^N
\]
denote the network output at time step \(t\), and let \(p(\mathbf{x}_t \mid \theta)\) be the predicted probability distribution parameterized by \(\theta\). Let \(q(\mathbf{x}_t)\) be the target distribution obtained from the dataset at time step \(t\), we can add  a temporal discount factor \(\gamma \in [0,1]\)  so that more recent time steps have a larger weight.

Then, the Kullback-Leibler (KL) divergence at time \(t\) is defined as:
\[
D_{KL}\Bigl( q(\mathbf{x}_t) \,\|\, p(\mathbf{x}_t \mid \theta) \Bigr)
= \sum_{i=1}^{N} q(x_{t,i}) \, \log \frac{q(x_{t,i})}{p(x_{t,i} \mid \theta)}.
\]

Thus, the overall loss function is given by:
\[
\mathcal{L}(\theta) = \sum_{t=1}^{T} \gamma^{t-1}\, D_{KL}\Bigl( q(\mathbf{x}_t) \,\|\, p(\mathbf{x}_t \mid \theta) \Bigr).
\]
where $|\gamma| < 1$ is a discounting factor to change the relative weights of prediction range.

In the special case where the target distribution is a delta distribution (e.g., represented as a one-hot encoded vector \( \mathbf{x}_t^{\text{target}} \)), the KL divergence reduces to the negative log-likelihood:
\[
\mathcal{L}(\theta) = -\sum_{t=1}^{T} \gamma^{\,t-1}\, \log p\Bigl(\mathbf{x}_t^{\text{target}} \mid \theta\Bigr).
\]
We have assumed a Gaussian distribution in all networks. 

\subsection{Inference}
The inference is then obtained by extracting from the estimated conditional probability distribution in a greedy inference pattern for the length of the prediction time. For each starting date of the forecast an ensemble of 50 forecast have been produced, in the following we are considering as a forecast the ensemble mean of each forecast. The analysis of the ensemble properties will be the subject of a forthcoming paper.

\section{Forecasting Monthly Mean Global SST}

The first case if the prediction of the global (limited to 60N - 60S)  monthly mean SST field. Fig. \ref{Fig2} shows the results of heuristic sensitivity analysis of \DS. The picture shows the performance of  forecast model made with \DS  compared  to Persistence and an operational  General Circulation Model (GCM) tha ti part of the Seasonal Forecasting Copernicus service, in this case those produced by CMCC (REF). Thefour panels, each of which examines the effect of a different hyperparameter or input configuration on forecast skill over lead times from Month 0 (M0) to Month 12 (M12). The vertical axis in each panel ranges from 0.5 to 1.0 is the spatial anomaly correlation coefficient of the patterns..

The top left panel compares different combinations of input variables, the lines correspond to different combination of the input fields a 
Sea Surface Temperature (SST) , Sea level Pressure (SP), Zonal Wind at 850mb (U850) and Top Outgoing Low Frequency Radiation (MTNLWRF). the results indicate that the best results is obtained using either SST and U850 or SST and SP. Keeping the combination of SST and U850, the top right panel shows the sensitivity to the hidden space dimension. The different configurations tested are 64, 128, 256, 512, and 1024. The best results is then carried over to the bottom left panel where the sensitivity to the number of lags considered is tested. This panel focuses on the impact of using different numbers of temporal lags as input. The configurations range from using six lags ([1, 2, 3, 4, 5, 6]) down to a single lag ([1]). Using shorter lags improve the performance on the short term, but degrades the performance on longer time scales, a better trade off is obtained using 4 lags, [1,2,3,4]. The final panel (bottom right)  explores how the depth of the network influences forecast skill by varying the number of hidden layers. Tested configurations include 1, 2, 4, 8, and 16 layers. The final best combination then involve using SST and U850, with a 256 hidden space, two transformer layer and 4 lags.

\begin{figure*}
 \centering
 \includegraphics[width=\textwidth]{ 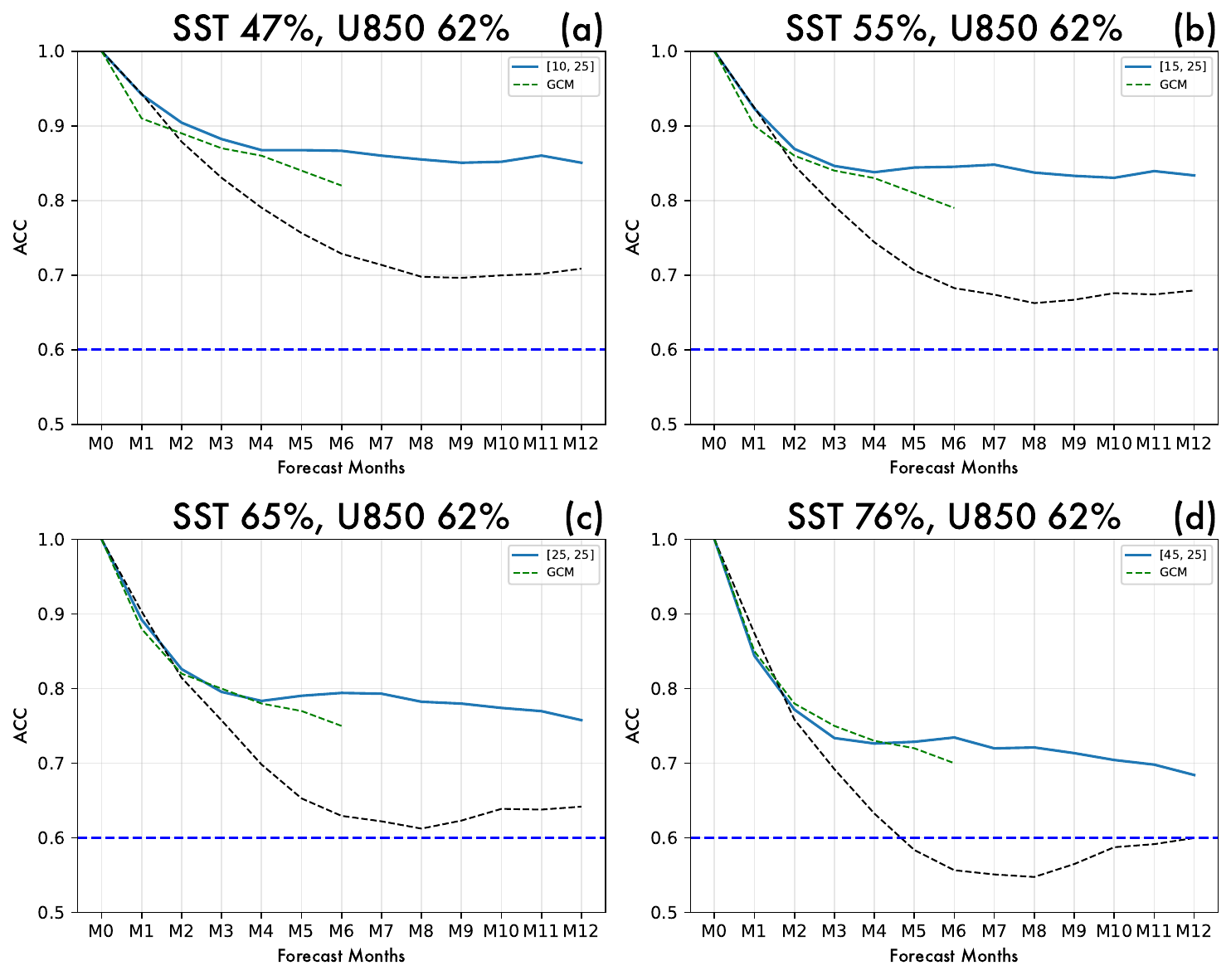}
 \caption{Sensitivity analysis of the \DS  forecasting system to the truncation level applied to the EOF (Empirical Orthogonal Function) decomposition of the SST input fields. The figure is arranged in four panels, each corresponding to  different pairs of truncation settings for SST and U850  corresponding to retaining 47\% of the variance (a), 55\% (b), 65\%(c), and 76\% (d), whereas the variance retained for the global U850 is kept fixed at 62\%.  In each panel, the x-axis represents forecast lead times from Month 0  to Month 12, while the y-axis displays the spatial anomaly correlation coefficient. The solid line illustrate the performance of the \DS model in comparison with benchmark forecasts from Persistence and an operational General Circulation Model (GCM) provided by CMCC as part of the Seasonal Forecasting Copernicus service, both projected on the same EOFs. This analysis reveals how varying the degree of EOF truncation affects forecast skill across different lead times. } 
\label{Fig3}
\end{figure*}

The Fig.\ref{Fig3} shows how \DS  reflects the changes in the retained EOFs in the input sequence. The EOFs have been computed on the Training period to avoid leaks in the future. Persistence and the GCM have been also projected and then reconstructed with the same EOFs. Predictability is different on different portion of the variance. In these experiments we have kept changed the number of EOF to represent different values of retained variance. respectively.  Keeping fewer EOFs gives an excellent performance compared to the Persistence and the GCM. The advantage becomes smaller as more and more EOFs are considered and the time behavior of the skill tend to follow a traditional drop with lead time and getting closer, i.e. no better or even worse, than Persistence. As the number of EOF is increased the skill of both \DS and the GCM are degraded, showing that the predictability is mostly in the large scale patterns that in any case dominate the variance. In the following we will show the result for the case where we have retained 55\% of the variance.

\begin{figure*}
 \centering
 \includegraphics[width=39pc]{ 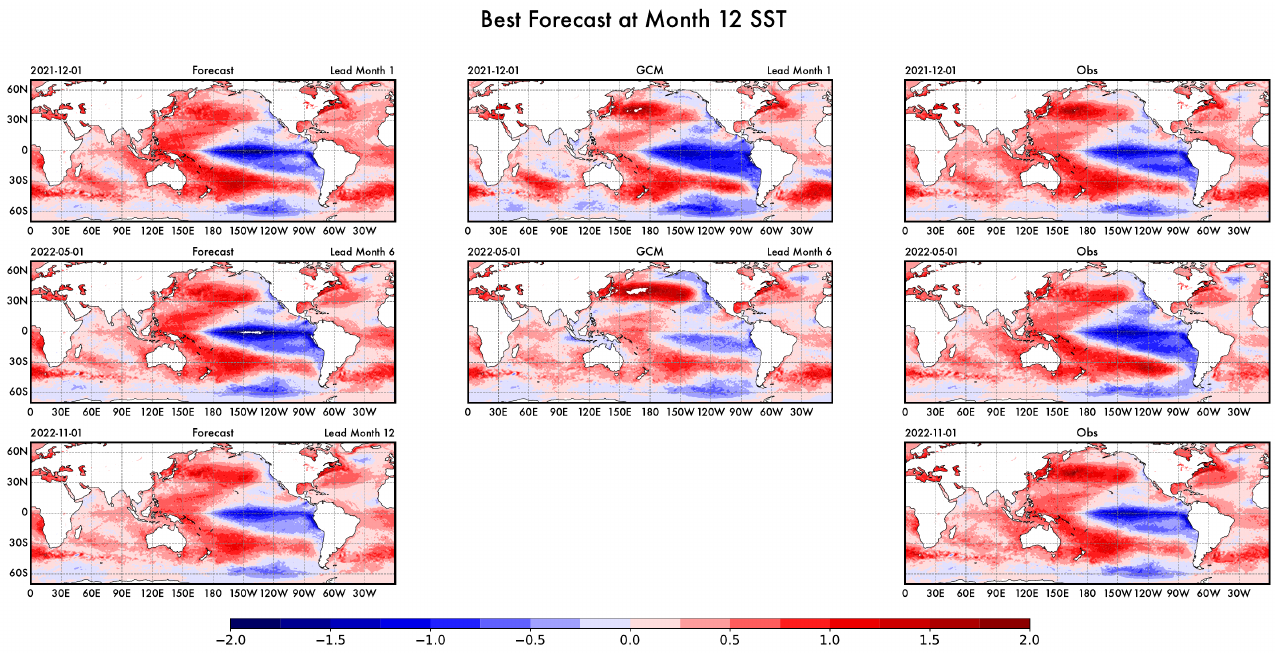}
 \caption{Comparison of Monthly Mean Sea Surface Temperature (SST) forecasts from the best-performing \DS configuration (15 SST EOF, 25 U850 EOF , temporal lags of 1-4,  hidden dimension of 256 one layer) and the best score among the initial dates at Month 12, against observationa and the operational GCM. The top row of panels (lead month 1) displays, from left to right, the \DS forecast, the GCM forecast, and the corresponding observations. The middle row (lead month 6) follows the same layout. The bottom row (lead month 12) presents the \DS forecast alongside the observed SST, omitting the GCM forecast for this lead time. Each map is showing the domain is for the global domain and between latitude (60S to 60N). The color scale (ranging from -2.0 to 2.0) denotes SST anomaly values in C.   } 
\label{Fig4}
\end{figure*}

To illustrate the performance of the system we can look at best forecast in ACC score at month 12, corresponding to the case with initial condition Nov 2021. It is shown in Fig.\ref{Fig4}. Here we show a comparison of Sea Surface Temperature (SST) forecasts from the best-performing \DS  configuration that uses 15 SST EOF against observations and the operational GCM. The top set of panels (initialized on 2021-12-01, lead month 1) displays, from left to right, the \DS forecast, the GCM forecast, and the corresponding observations. The middle set (valid for lead month 6) follows the same layout. The bottom set (lead month 12) presents the \DS forecast alongside the observed SST, omitting the GCM forecast for this lead time that is not available since the GCM forecasts extends only up to six months. The domain extension is for the global domain between  latitude (60S to 60N), and the color scale (ranging from -2.0 to 2.0) denotes SST anomaly values in C.  The \DS system makes an exceptional performance maintaining the La Ni\~na conditions that persists for along time, whereas the GCM looses them at month 6. Also the weak Indian Ocean dipole structure  is captured correctly and the Gulf Stream front is more marked than in the GCM.

\begin{figure*}
 \centering
 \includegraphics[width=\textwidth]{ 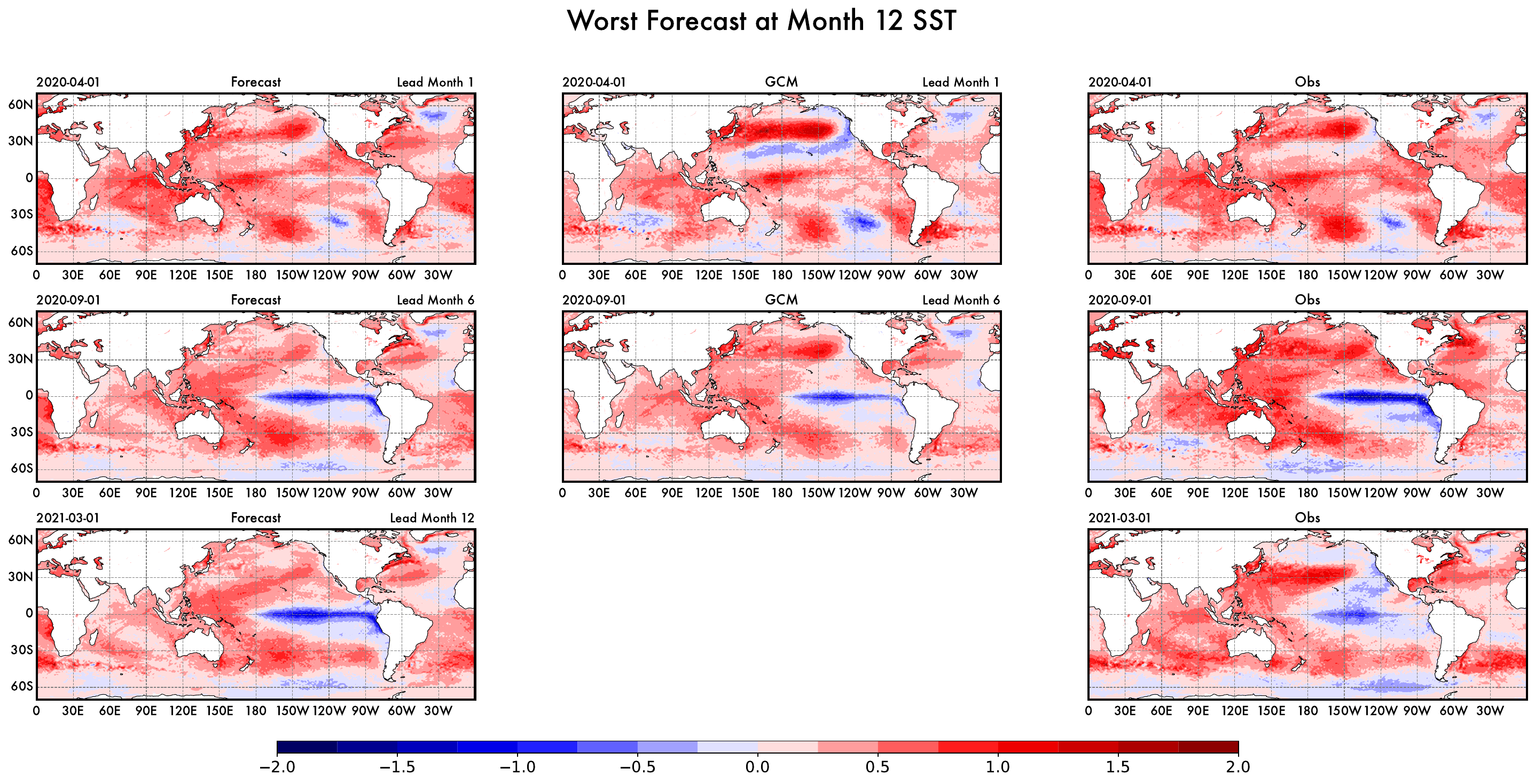}
 \caption{As in Fig.\ref{Fig4}, but for the worst forecast at Month 12.}
\label{Fig5}
\end{figure*}

On the other hand, Fig.\ref{Fig5}, shows the worst forecast, with the initial condition March 2020. The development of the cold conditions in the tropical Pacific is insufficient and the intensity of the warm anomalies in the West Pacific is too weak.  The cold anomalies in the Southern Ocean are also missing. 

\begin{figure*}
 \centering
 \includegraphics[width=\textwidth]{ 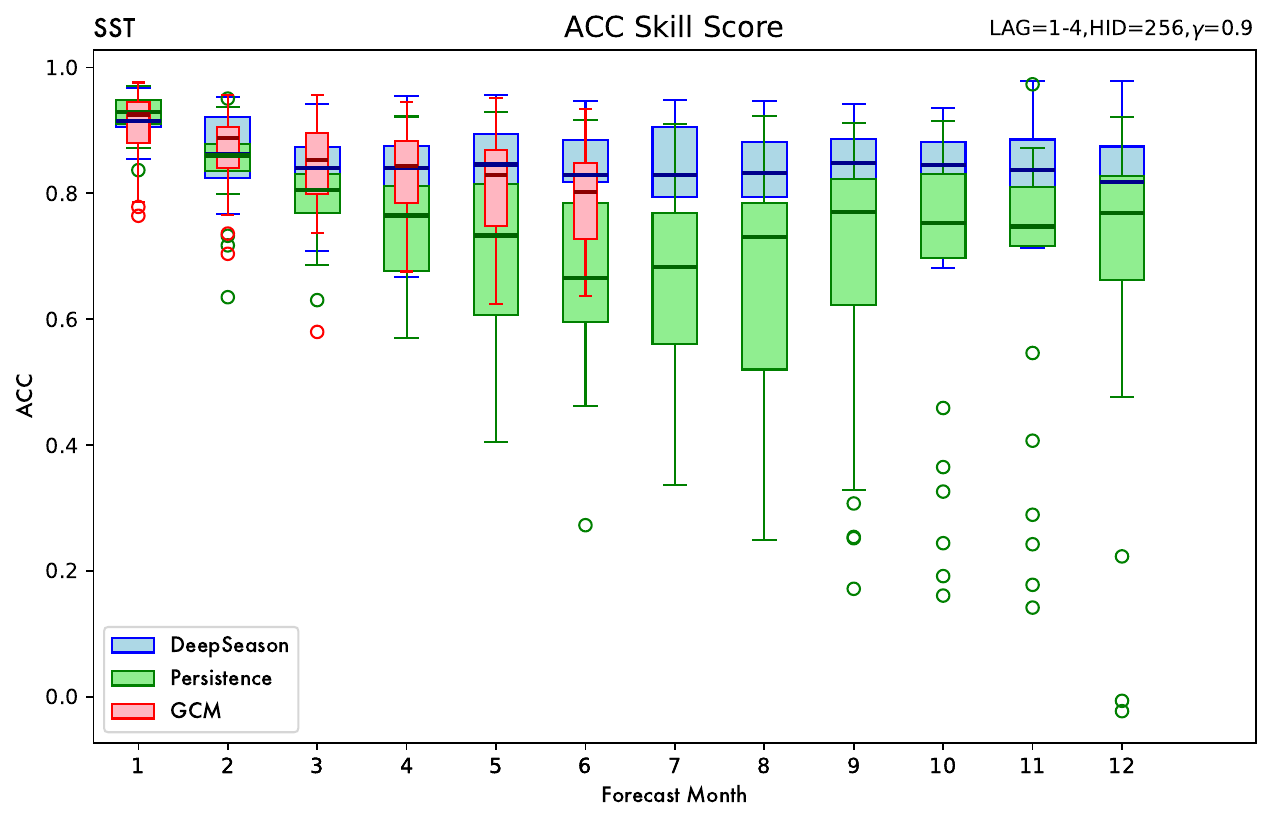}
 \caption{Boxplot of spatial anomaly correlation coefficients (ACC) for SST forecasts for the initial dates from produced by the \DS model versus Persistence and the operational GCM. The forecasts are generated using a \DS configuration with (SST,U850) as input, temporal lags of 1 - 4, and a hidden dimension of 256.  The boxplot summarizes the distribution of forecast skill across lead times. For each forecast month, the boxplot displays the median (central line), the interquartile range (the box edges), and the full range of the data (whiskers). Lightblue is used for the \DS forecast, lightgreen for the persistence forecast, and lightpink for the GCM-based forecast. Median and quartile values are shown in dark blue, dark green, and dark red, respectively. }
 \label{Fig6}
 \end{figure*}

A comprehensive look of the behavior of the entire set of forecasts can be seen in Fig. \ref{Fig6} showing the correlation skill score as a box plot for the initial starting dates from Dec 2019 to Dec 2021. Also shown are the scores for Persistence and the operational GCM. The forecasts are generated using a \DS configuration with (SST,U850) as input, temporal lags of 1-4, and a hidden dimension of 256.  The plot summarizes the distribution of forecast skill across several lead times. For each forecast month, the boxplot displays the median (central line), the interquartile range (the box edges), and the full range of the data (whiskers). Lightblue is used for the \DS forecast, lightgreen for the persistence forecast, and lightpink for the GCM-based forecast. Median and quartile values are shown in dark blue, dark green, and dark red, respectively. \DS has a performance comparable to the GCM especially in the early part of the forecast and it is better at month four and five. It is always equal or better than the persistence maintaining a good level of performance also at later lead times. Furthermore, it has few outliers and a relatively stable performance.

\begin{figure*}
  \centering
  \includegraphics[width=\textwidth]{ 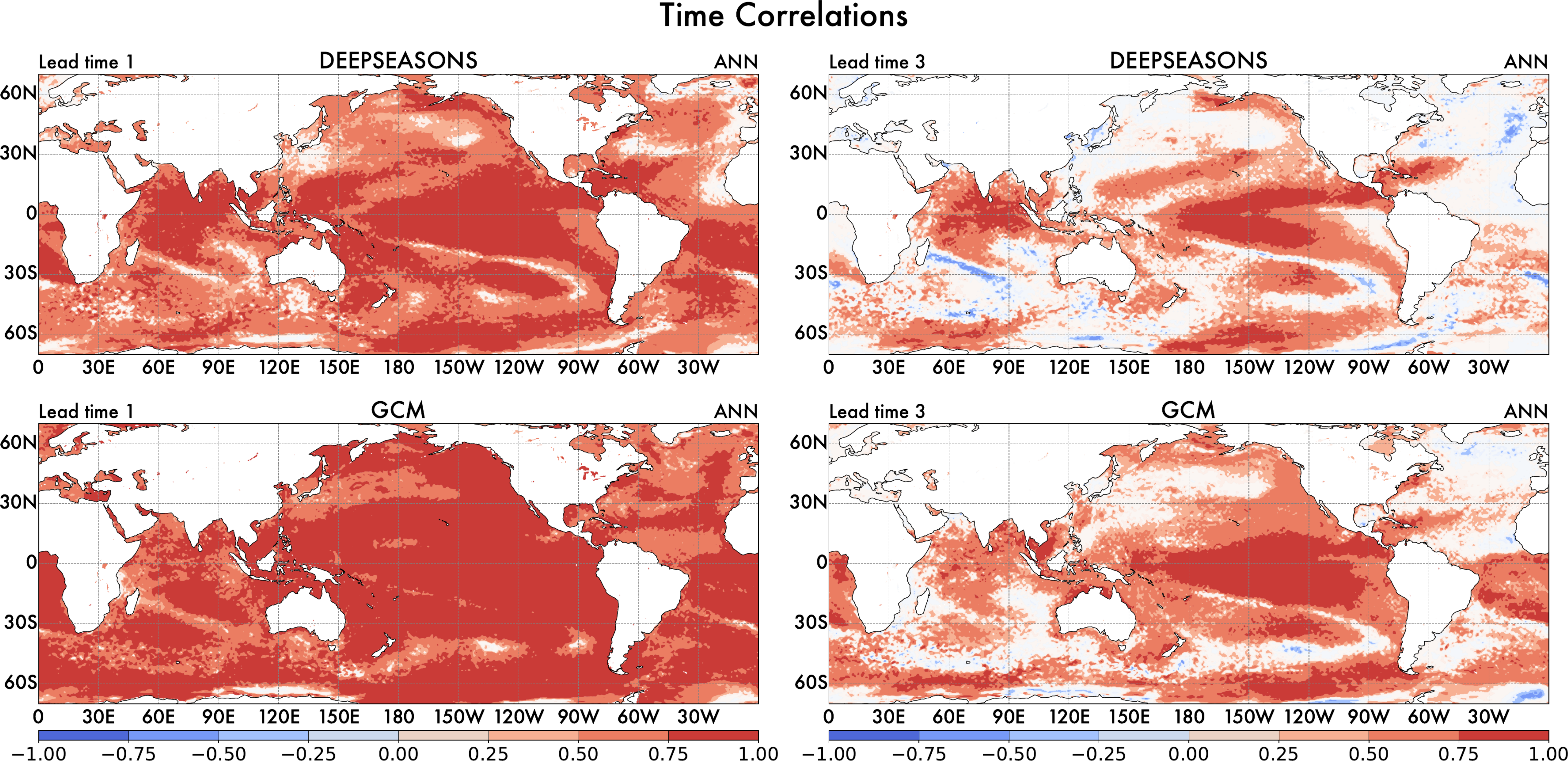}
  \caption{Spatial maps of time correlations for forecasts  computed at a forecast lead time of one and three months. The correlations are computed over the 28 forecasts of the test period. The top panels display the \DS model results, while the bottom panels showsthe corresponding correlations from the operational GCM. Dark colors indicate values significant at 10\%.}
  \label{Fig:SST-Time-Corr}
\end{figure*}

Fig.\ref{Fig:SST-Time-Corr} is showing the time correlation of the forecasts with \DS compared with the operational GCM. These correlations were computed by correlating the forecast at specific lead times with the corresponding observations at the same lead times point by point. He give us an appreciation of how consistent the forecasts are with  observations on the entire set of forecasts. The left column shows the results at the lead time one. The correlations are very high both for \DS and the  GCM results. They're well above .75 over most of the global ocean with some limited bad results over the polar region mostly in both Hemispheres.  \DS shows somewhat less favorable results in the central Pacific and over the north Atlantic Ocean. The right column shows the same quantity for lead time three. We noticed a significant degradation of the skill of the forecast, both for \DS and the GCM, with  most of the skill concentrated in the tropical Pacific, Pacific Ocean and Indian Ocean. The skill becomes worse in the northern latitudes both of the Pacific and Atlantic oceans but it is possible to note that \DS has better results in the Indian Ocean and south of Australia in the high south latitudes of the Indian Ocean. \DS has  persistent difficulties in the northern Pacific and Atlantic oceans. 

The following figure  (Fig.\ref{Fig:SST-Time-Corr-2}), is showing similar results for different lead times. The left column is showing lead times at month six for both \DS and the GCM. The loss of skill continues as we progress into the forecast, but we notice that in general \DS is quite comparable with the skill of the GCM and occasionally it is better, like in the Indian Ocean, but it can be also worse, as for instance in the southern Atlantic. The correlations for month nine on the right column is available only for \DS as the GCM forecasts are not available beyond the month six. The skill that is spotty and again is the concentrated in the tropic equatorial Pacific and Indian Ocean and it is becoming much more noisy elsewhere.

\begin{figure*}
  \centering
  \includegraphics[width=\textwidth]{ 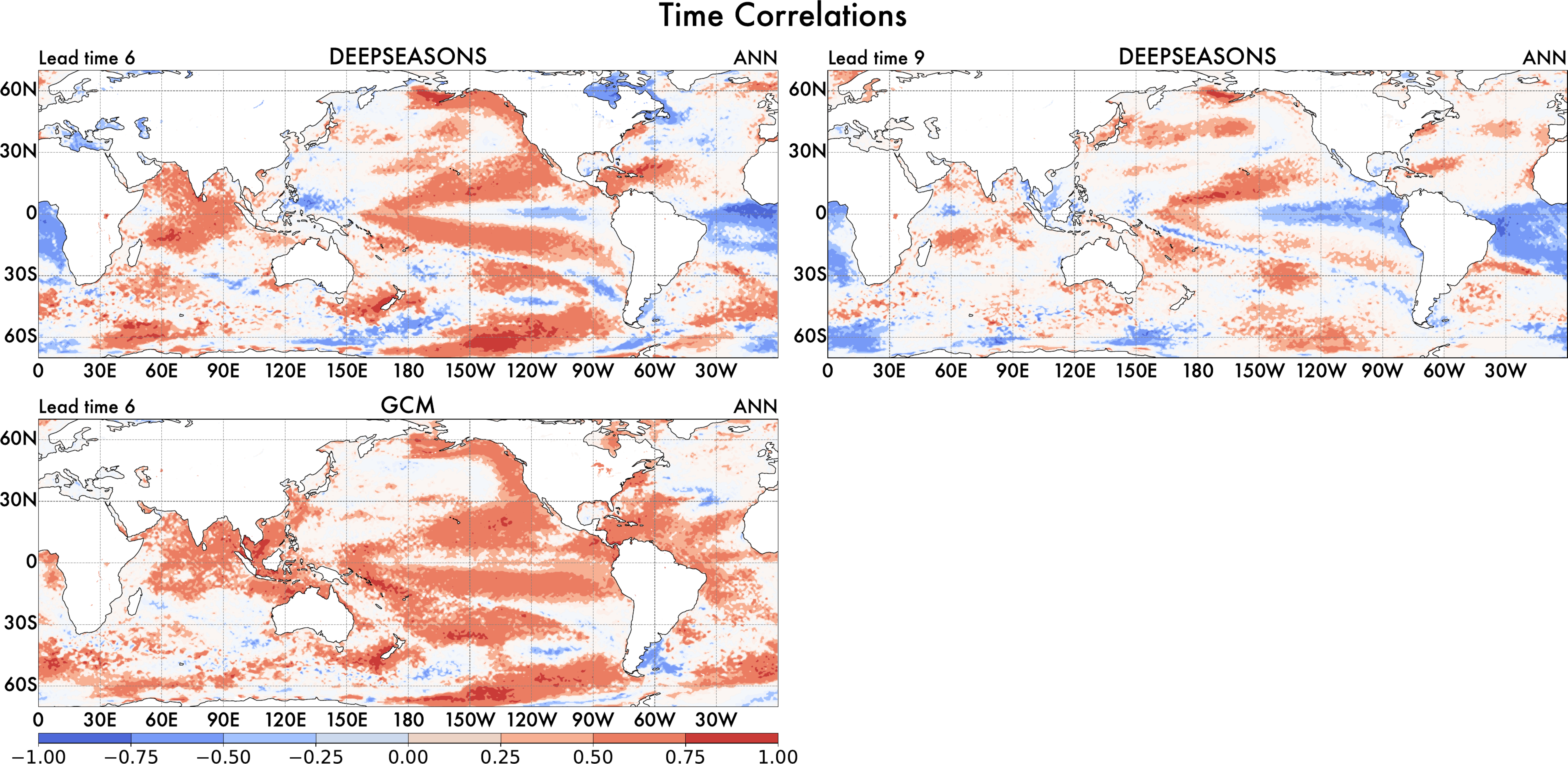}
  \caption{As in Fig.\ref{Fig:SST-Time-Corr} but for lead times six and nine. There are no GCM forecasts for months nine. }
  \label{Fig:SST-Time-Corr-2}
\end{figure*}

Another way to analyze the performance of the model can be seen in Fig.\ref{Fig:SST-Time-Corr}, showing the point-by-point time correlation between forecasts at specific lead times and the ERA5 verification. In the top panel, \DS  forecasts reveal large areas of high positive correlation in the Indian Ocean, the tropical Pacific and Atlantic Ocean. and less positive correspondence in the East Equatorial Pacific Ocean. Overall, the results indicates that the \DS  tends to maintain relatively high correlation values over broad regions, suggesting a good ability to capture the temporal evolution of SST anomalies. 
The bottom panel, which shows the forecast from the operational GCM, is showing a very similar distribution of time correlations. Some differences are noticeable, especially in the equatorial East Pacific, where the GCM seems to be able to capture better the variability there, on the other hand, \DS is better in the North and equatorial Atlantic. 

Figure \ref{Fig:SST-Time-Corr-2} displays the same quantity for lead times six and nine. In the last case, we lack the GCM forecasts. In this scenario, we observe a generally consistent pattern of skill between the DS and the GCM, with some variations in the Atlantic Ocean, where the GCM appears to be more accurate, and in the Indian Ocean, where the DS yields better results. The predictability significantly deteriorates by month nine (left panel), although some residual areas of skill remain visible. The interpretation of this kind of picture is somewhat challenging. We adopted a restricted view, considering only positive values of correlation as indicators of skill. However, it is common for large negative values to also appear. 

Overall, the contrast between the two panels highlights the capability of the  \DS   to maintaining good forecast skill over a six-month lead time, on the same par, sometimes better than performance of the operational GCM. considering the difference in costs and effort, is a remarkable result. 

\section{Forecasting Monthly Mean Temperature at 2m}

\subsection{Europe}

In this section, we exploit the property of \DS to be tailored for specific regions in an important case: forecasting near-surface temperature, nominally at 2 meters, T2M, over a specific region, in this case the European region. Temperature at 2m is a key metric for a range of societal and economic decisions, from energy demand forecasting to agricultural planning. Europe, with its diverse climate zones and high-quality observational datasets, presents a compelling testbed for this methodology. 

\begin{figure*}
    \centering
    \includegraphics[width=\textwidth]{ 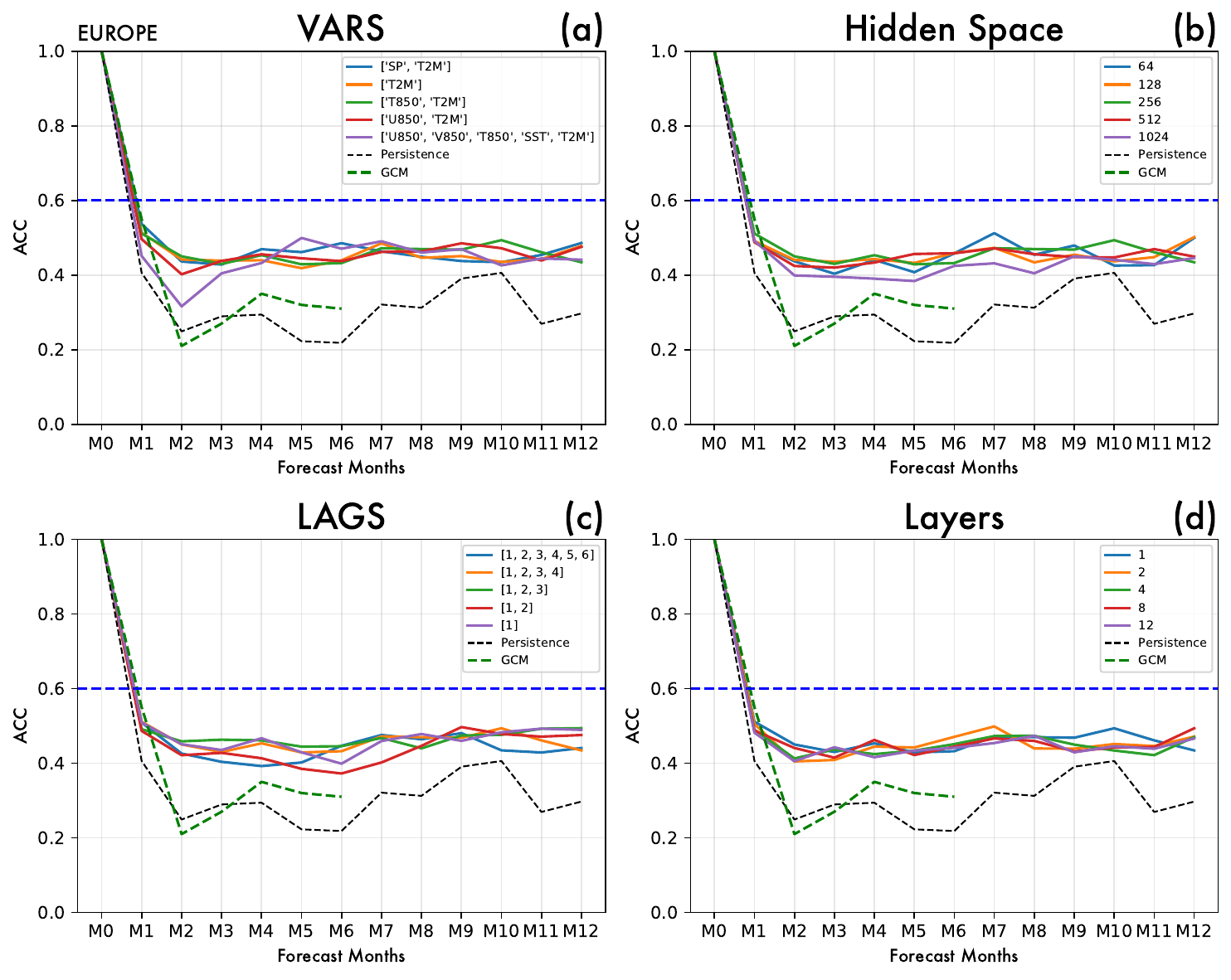}
    \caption{Heuristic sensitivity analysis of \DS for T2M on the European region. Each panel compares the mean forecast skill (spatial anomaly correlation, 0.5-1.0) of the \DS based model versus Persistence and a CMCC GCM over lead times from Month 0 (M0) to Month 12 (M12) for the set of forecasts in the Test set. The top-left panel (a) examines the input variable combinations  with the best results obtained using either T2M or T2M and T850. Using T2m and T850, the top-right panel (b) explores the impact of the hidden space dimension (64, 128, 256, 512, and 1024). The bottom-left panel (c) investigates the sensitivity to the number of temporal lags, showing that while shorter lags boost short term performance, a tradeoff is achieved with four lags (1,2,3,4). Finally, the bottom-right panel (d) evaluates the influence of network depth (1, 2, 4, 8, and 16 hidden layers), with the optimal configuration being a single transformer layer.}
    \label{fig:T2M-forecast}
\end{figure*}

Figure \ref{fig:T2M-forecast} presents  the sensitivity of model performance across different design and configuration parameters for the temperature forecasts. The skill score used here is again the anomaly correlation coefficient over the region (ACC) between the forecast  and the verification are also monthly mean ERA5 data, projected onto the same EOF used a basis for the model. The vertical axis across all panels represents the skill score, while the horizontal axis tracks forecast lead time in months (M0 to M12). Each subplot investigates a different aspect of model sensitivity, comparing performance against baseline persistence and GCM benchmarks. The dashed black line is the skill for the persistence forecast, whereas the green dashed one is the result for the operational GCM.

In the top-left panel, the analysis focuses on the impact of variable selection on forecast skill. Multiple variable combinations are evaluated, including pairs like Sea level Pressure and Temperature (SP, T2M), or pressure levels and T2M  (T850, T2M), as well as a more comprehensive set including U850, V850, T850, SST, and T2M. The target region is the European area (30N to 70N and 20W to 50E)), whereas the region for the other variables is global. All input variables are truncated to  a number of EOF representing about 75\% of explained variance. over the respective regions. The EOF have been computed over the European region, except for the case of the SST where we are using a global domain to try to capture some of the remote connections.

The top-right panel explores how the choice of discount factor affects performance, using the best-performing model configuration identified from the variable selection experiment. Discount factors ranging from 0.1 to 1.0 are tested, where higher values imply slower decay of the weight of future targets.  Results show that moderate discounting (e.g., 0.7 to 0.9) offers the most consistent gains, especially between months 2 to 8, indicating that appropriately weighting the loss function can enhance the temporal learning dynamics of the model.

The bottom two panels deal with  the sensitivity of the structural model. The bottom-left panel examines the impact of the number of Empirical Orthogonal Function (EOF) components employed in preprocessing on the best model resulting from the preceding analysis. While increasing the EOF count from 75\% to approximately 90\% of variance explained does not significantly enhance skill at longer lead times, the bottom-right panel reveals that varying the number of neural network layers has a notable effect. Deeper architectures underperform, suggesting that shallower designs yield satisfactory results. These findings collectively emphasize the importance of striking a balance between model complexity and generalization.  To finalize the analysis we chose a  final configuration  then involve using  regional T2M and T at 850mb with five EOF, discount at 0.9, one transformer layer in both encoder and decoder and a hidden dimension of 256, even if, considering Table \ref{Table-1}, other choices of variables are also possible.

\begin{table*}[h]
\caption{RMS errors over the European area for the T2M temperature for various forecast made with \DS and different input variables. The minimum values are emphasized in red. Also indicated are the error for persistence (PERS) and the operational model (GCM).}
\begin{center}
\begin{tabular*}{\textwidth}{@{\extracolsep{\fill}}lccccccccccccc@{}}

VARS & M0 & M1 & M2 & M3 & M4 & M5 & M6 & M7 & M8 & M9 & M10 & M11 & M12 \\

SP, T2M & 0.00 & \textcolor{red}{0.61} & 0.79 & 0.79 & \textcolor{red}{0.74} & 0.76 & \textcolor{red}{0.72} & 0.72 & 0.72 & 0.73 & 0.71 & 0.69 & 0.67 \\
T2M & 0.00 & 0.65 & 0.79 & \textcolor{red}{0.74} & 0.76 & 0.77 & 0.77 & \textcolor{red}{0.70} & 0.74 & 0.74 & 0.72 & 0.70 & \textcolor{red}{0.66} \\
T850 T2M & 0.00 & 0.62 & \textcolor{red}{0.73} & 0.75 & 0.74 & 0.77 & 0.76 & 0.75 & \textcolor{red}{0.70} & 0.71 & \textcolor{red}{0.64} & \textcolor{red}{0.63} & 0.70 \\
U850, T2M & 0.00 & 0.66 & 0.81 & 0.79 & 0.77 & 0.77 & 0.78 & 0.71 & 0.70 & \textcolor{red}{0.69} & 0.67 & 0.71 & 0.66 \\
U850, V850, T850, SST, T2M & 0.00 & 0.70 & 0.92 & 0.79 & 0.79 & \textcolor{red}{0.71} & 0.73 & 0.70 & 0.71 & 0.71 & 0.74 & 0.68 & 0.68 \\
PERS & 0.00 & 0.89 & 1.32 & 1.30 & 1.34 & 1.48 & 1.42 & 1.20 & 1.20 & 1.08 & 1.02 & 1.26 & 1.30 \\
GCM & 0.00 & 0.86 & 0.99 & 0.91 & 0.81 & 0.86 & 0.84 &  &  &  &  &  &  \\

\end{tabular*}
\end{center}
\label{Table-1}
\end{table*}%

The correlation factors over a small area may be  too sensitive to the details of the forecast. The Mean Root Square Error shown in Table \ref{Table-1} may give a different angle on the performance. \DS yields a lower error than the persistence in every instance and it is lower than the GCM also. Using only the T2M temperature is giving already good results, they are improved adding the temperature at 850 and the sea level Pressure, but adding more input features does not result in significant improvement in the simple design used by \DS. Analysis of the RMSE in the other experiments confirm the choice of the combination of parameters already adopted.

\begin{figure*}[h]
    \centering
    \includegraphics[width=\textwidth]{ 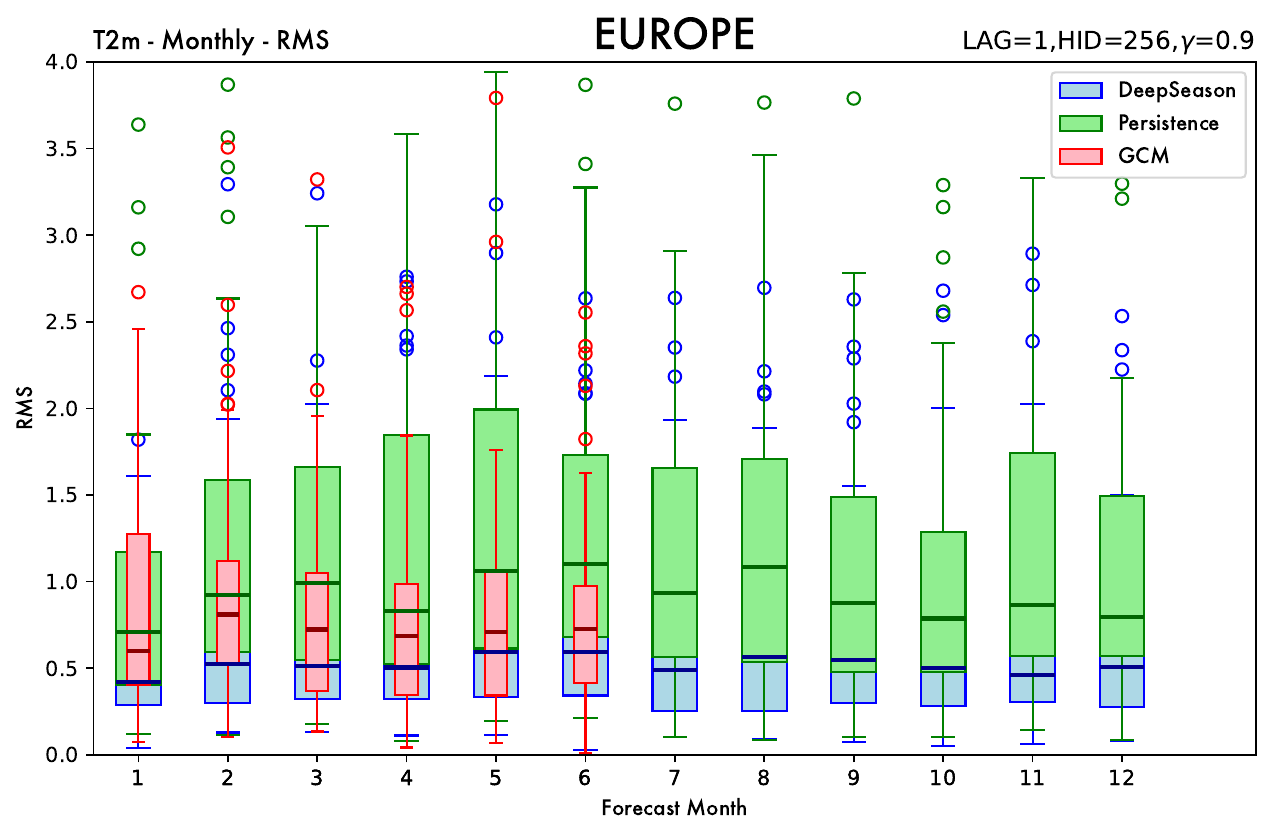}
    \caption{Root Mean Square Error box plot for the verification field T2M and for the European region. 
    Lightblue is used for the model forecast, lightgreen for the persistence forecast, and lightpink for the GCM-based forecast. Median and quartile values are shown in dark blue, dark green, and dark red, respectively. The input fields used are T850 and 'T2M, using 75\% of the variance,  the lags considered are four.  The size of the hidden space of the model is 256 and the discount parameter for the loss function is 0.9.}
\label{Fig:boxT2M}
\end{figure*}

 Fig. \ref{Fig:boxT2M} shows the  box plot for the RMSE displaying the distribution of errors for the starting dates from  2017-06-01 to 2021-12-01 for a total 55 forecast for T2M over the EUROPE region. The sensitivity to the length of the input sequence was not investigated exhaustively, but a limited set of experiments showed that the best results are obtained with the sequence of 18 months that is being used here. Lightblue is used for the model forecast, lightgreen for the persistence forecast, and lightpink for the GCM-based forecast. Median and quartile values are shown in dark blue, dark green, and dark red, respectively. The input fields used are T850, T2M.  In this case enoth EOFs to explain 75\% of the variance are retained,  the size of the hidden space of the model is 256 and the discount parameter for the loss function is 0.9. 
 
\begin{figure*}
 \centering
 \includegraphics[width=39pc]{ 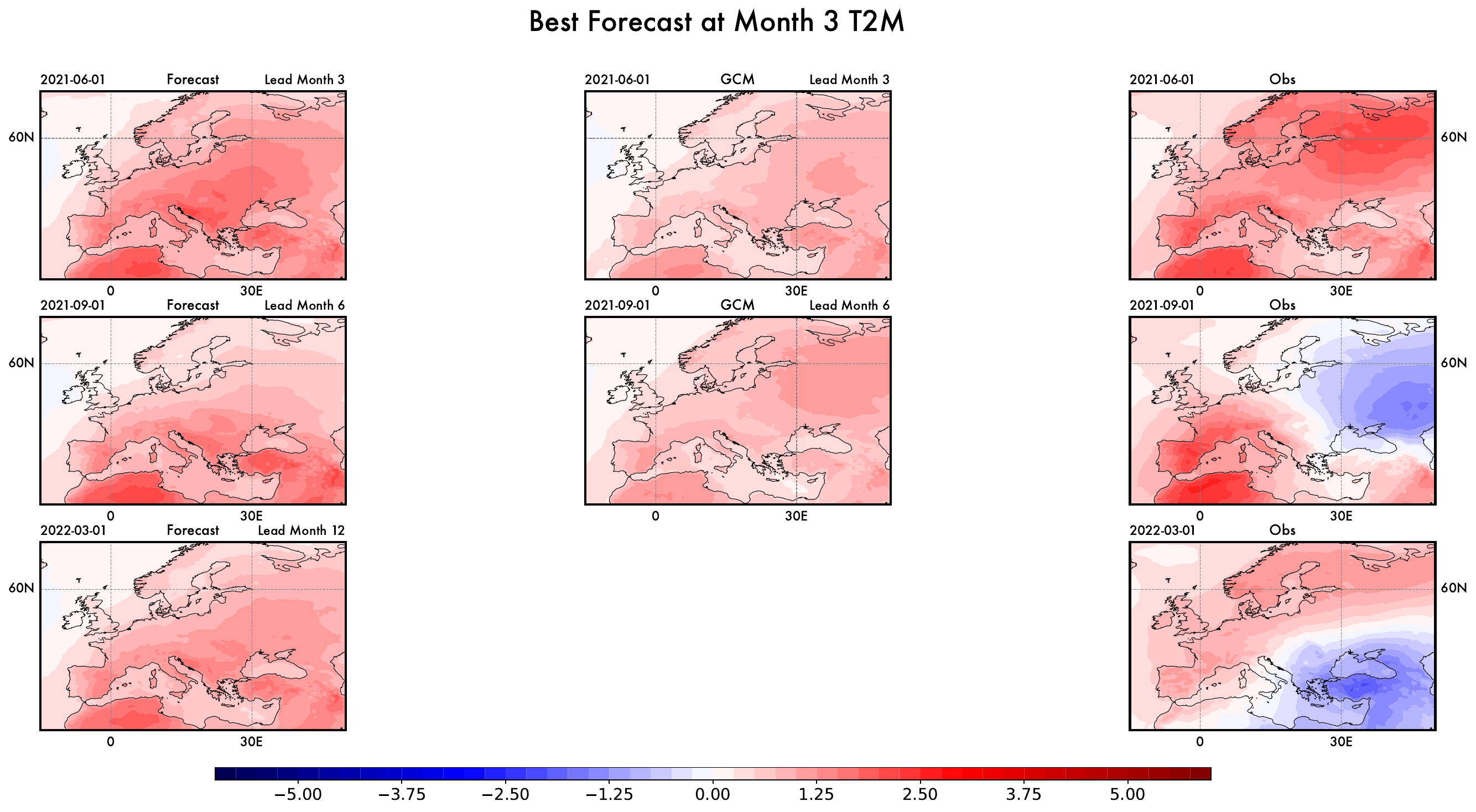}
 \caption{Comparison of anomaly temperature at 2m forecasts from the best-performing \DS configuration (5 T2M EOF, 5 T850 EOF , temporal lags of 1-4,  hidden dimension of 256 and  one transformer layer) and the best score in terms of minimum RMSE among the initial dates at Month 3, against observations and the operational GCM. The top row of panels (lead month 3) displays, from left to right, the \DS forecast, the GCM forecast, and the corresponding observations. The middle row (lead month 6) follows the same layout. The bottom row (lead month 9) presents the \DS forecast alongside the observed T2M, omitting the GCM forecast for this lead time that is not available. Each map is showing the domain is for the global domain and between latitude (60S to 60N). The color scale (ranging from -5 to 5 ) denotes T2M anomaly values in Celsius. } 
\label{Fig:BestT2M}
\end{figure*}

The best forecast of \DS (minimum RMSE) at month three is shown in \ref{Fig:BestT2M}. \DS is successful in capturing the  general pattern and amplitude of the anomaly at month three, but it fails to describe the  intensification of the warming over the East Mediterranean in the following months. On the other hand, the GCM is weakening prematurely at month six. In general , we can note in \DS a tendency toward an underestimation of the monthly variability and toward slower monthly variations. This tendency is shown also in other individual forecasts (not shown). 

\begin{figure*}
 \centering
 \includegraphics[width=39pc]{ 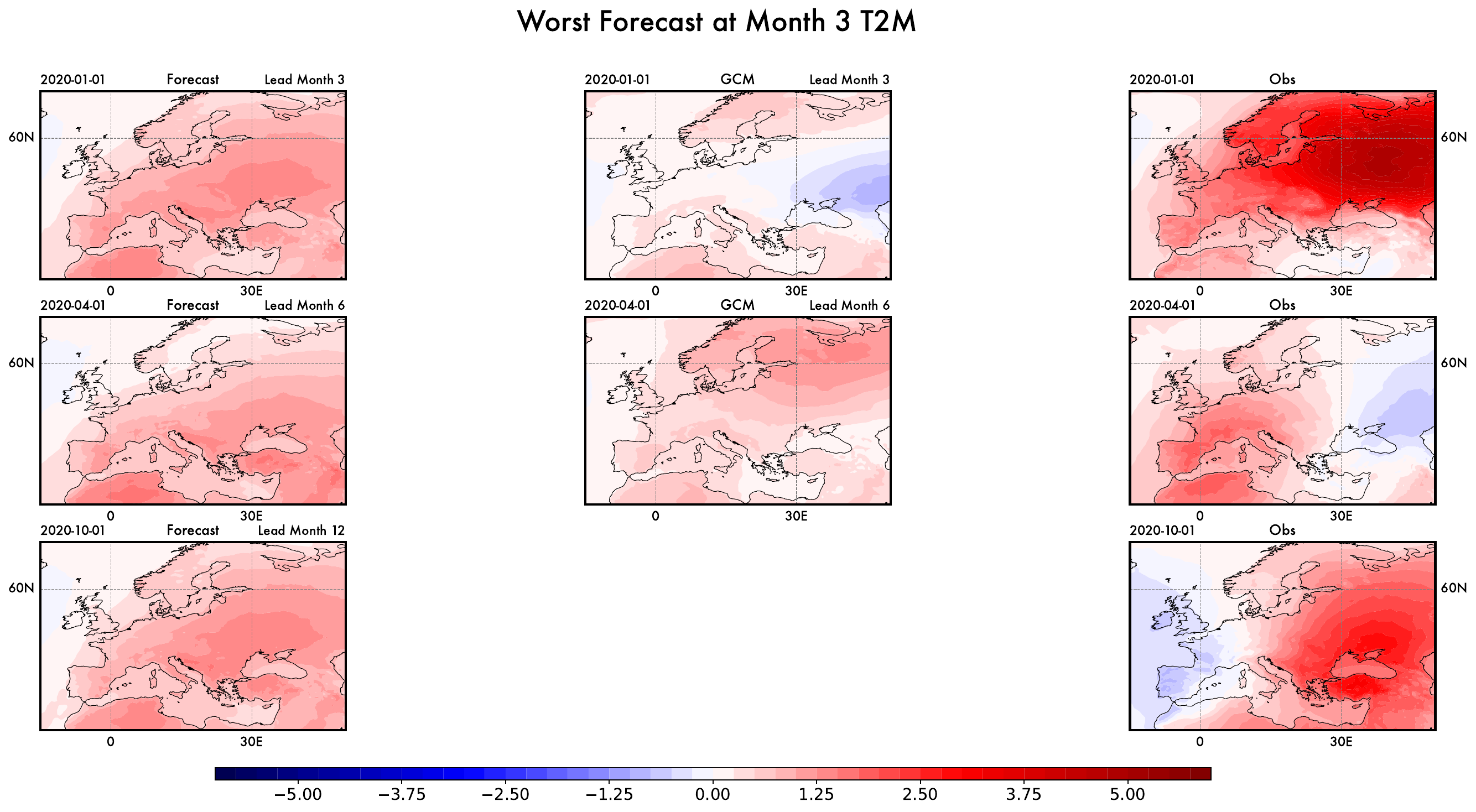}
 \caption{As in Fig.\ref{Fig:BestT2M} but for the worst forecast at month three } 
\label{Fig:WorstT2M}
\end{figure*}

The worst forecast (maximum RMSE) at month three is shown in Fig.\ref{Fig:WorstT2M}. We can also see here that the main source of error for \DS is a tendency to underrepresent contrasting pattern of warm and cold anomalies.

On the other hand, the worst forecast in Fig.\ref{Fig:WorstT2M} shows probably the main shortcoming of \DS: the fact that it tends to underestimate variability from month to month and to generate variations that go in general slower and smoother than the observations. In this forecast, for instance, at month three \DS forecast is already missing to capture the strong gradient in anomalies over eastern Europe and Black Sea, and it is sort of locked into a pattern of warm anomalies over the south Mediterranean and of relatively cooler over Northern Europe. \DS then evolves this pattern with slow changes that do not reflect the fast variation that observations show. However, the GCM is showing some similar problems. It is capturing the pattern at month three better than \DS, but the amplitude of the anomaly is in general underestimated. It also evolves towards the general warmer pattern that prevails in models six and nine, but in six, the amplitude is well underestimated by contrast \DS at month  six is already having a better representation of the intensified warming over south Europe, and the Mediterranean.

\begin{figure*}
 \centering
 \includegraphics[width=39pc]{ 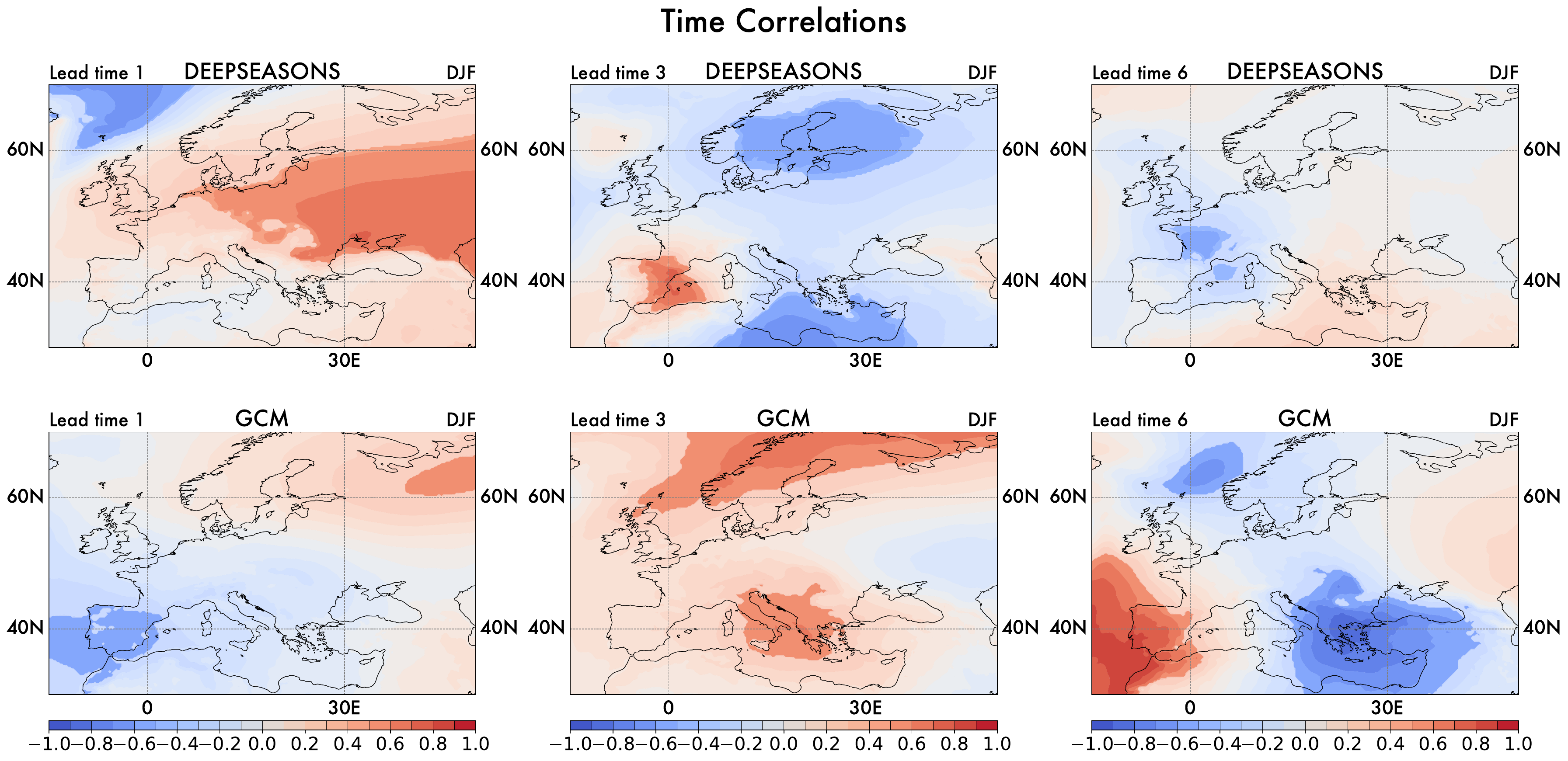}
 \caption{Time Correlation of Monthly Mean 2-meter Temperature Forecasts  for DJF Initial Conditions. The correlation is computed by considering all months at a specific lead time with the corresponding observations in ERA5. This figure illustrates the correlation  as predicted by the \DS  and the General Circulation Model (GCM) at various lead times for initial conditions set during the December-January-February (DJF) period. More saturated colors indicate areas that are significant at the 10\% level. The top three panels display the \DS time correlation maps at lead times of  3, 6, and 9 months, respectively. The lower  panels show the corresponding time correlations obtained from the GCM for the same lead times. 
These correlation maps are crucial for evaluating the models' performance in predicting temperature anomalies over varying forecast horizons. The GCM forecasts are available only up to lead time six.} 
\label{Fig:Time-Corr-T2M}
\end{figure*}

\begin{figure*}
 \centering
 \includegraphics[width=39pc]{ 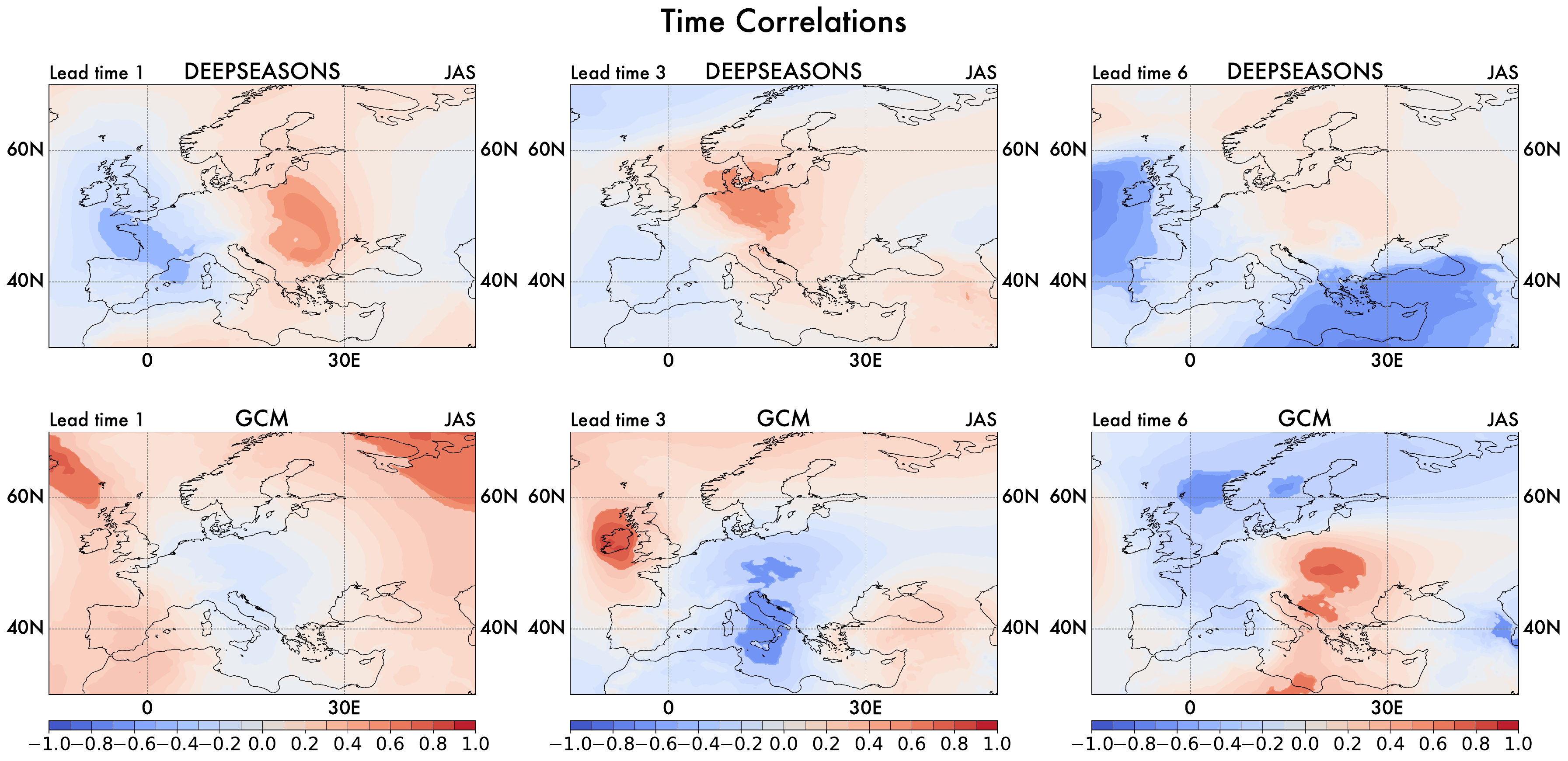}
 \caption{As in Fig.\ref{Fig:Time-Corr-T2M}, but for JAS initial condition} 
\label{Fig:Time-Corr-T2M-JAS}
\end{figure*}

In order to get a more coherent evaluation of the forecast performance is better to look at some quantity that gives a more comprehensive view of the overall performance of the model. Such a quantity is the correlation  in time with observations at the same point, as we have shown already in Fig.\ref{Fig:SST-Time-Corr} for the global SST. Figure\ref{Fig:Time-Corr-T2M}  is showing the correlation in time at different lead times both for \DS and the operational GCM in boreal Winter (DJF).  We have here 55 forecast available and so we can look separately at different seasons. The top row is showing the time correlation and lead time one, three and six months for \DS, the bottom row is showing the same quantity for the operational GCM.  Taking into consideration only the positive correlation as representative of real predictive skill we can evaluate that \DS is capable of giving a good forecast  at lead time of  one month, over most of the continental part of Europe and Western Russia.  This capacity is stronger than the operational GCM at the same time, but it  gets weaker when  you go to Month three where \DS is showing predictive capacity  only over western Europe.  The GCM is doing better at this time lead showing a more distributed skill that extends over northern Europe and the Mediterranean area.  However, such skill is lost by month six and is mostly limited to the Iberian peninsula. At the same time, \DS is maintaining predictive skill over the South of the Mediterranean Sea and some skill is also present over Western Europe and the British islands.  

Fig.\ref{Fig:Time-Corr-T2M-JAS} shows the a similar time correlation during the Summer (JAS) season. Predicting monthly mean forecasts in this period presents considerable challenges for both \DS and the GCM models and small areas of positive correlation can be seen over Eastern Europe.
The interpretation of these correlations, which is a standard practice in the evaluation of seasonal and longer-term forecasts, raises several key issues. One critical concern is the understanding of strong negative correlations that indicate a tendency for the model to consistently predict opposite anomalies compared to observations, which could indicate a systematic bias or model deficiencies.

\subsection{North America}

We are now considering the case of North America. We are here considering a region defines by  latitudes  boundaries (25N,70N), and longitudinal boundaries of 200E,310E, essentially the whole of the North American continent.  
The regions considered in this case would allow for the expected remote effect of the tropical Pacific, therefore the SST and pressure input variables were considered over the tropics, in the latitudinal band 35N-35S, whereas the temperature at 850 was used over the same area of the target 2m temperatures. We will select the case using the SST as additional input variables together of course with T2M over the target region.

\begin{figure*}[h]
\centering
\includegraphics[width=\textwidth]{ 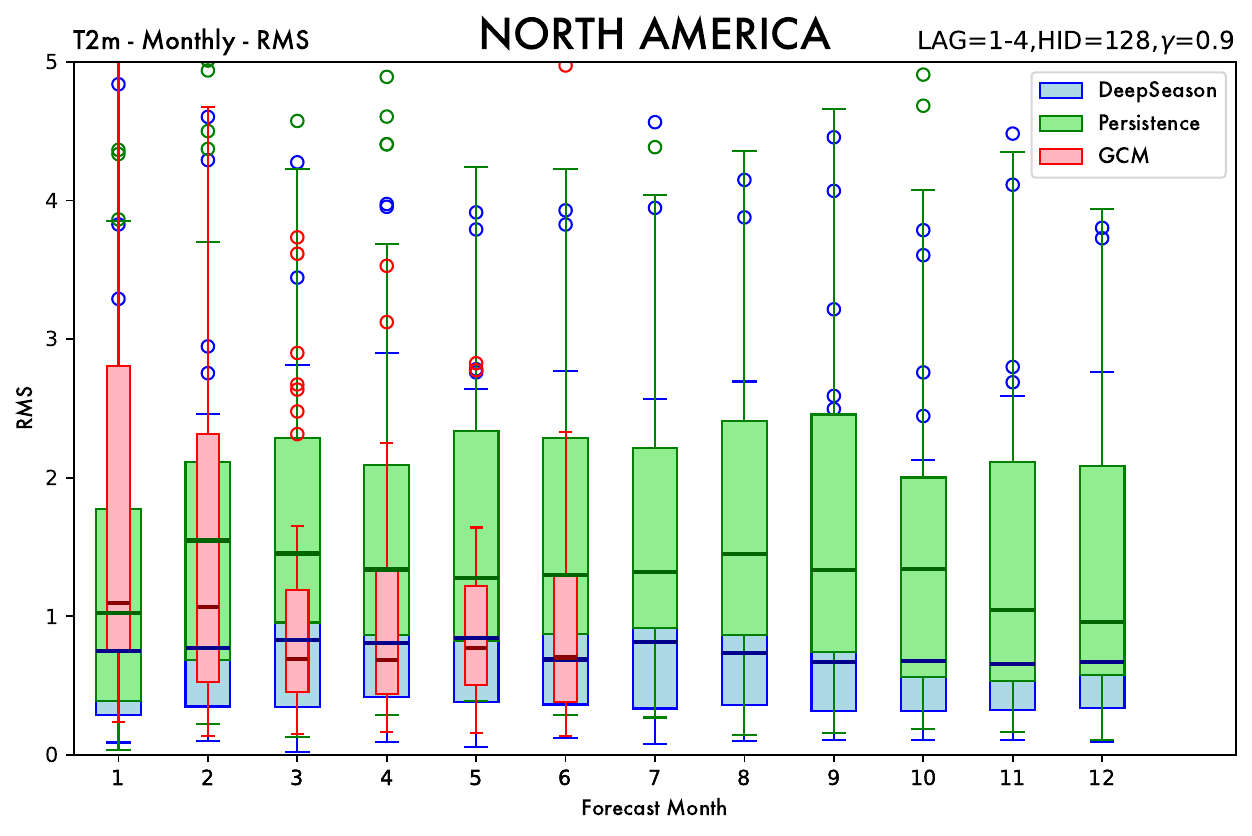}
\caption{RMSE box plot for the verification field T2M and for the North America region.  Lightblue is used for the model forecast, lightgreen for the persistence forecast, and lightpink for the GCM-based forecast. Median values are shown in dark blue, dark green, and dark red, respectively.}
\label{Fig:BoxNAT2M}
\end{figure*}

%
Fig.\ref{Fig:BoxNAT2M} shows the RMSE box plot for T2M over  the North America region. Light blue represents the model forecast, light green the persistence forecast, and light pink the GCM-based forecast. Median are shown in dark blue, dark green, and dark red, respectively. As in the European case, the input fields used are T850 and T2M. In this case, the EOF truncation involves retaining enough EOF to explain approximately 75\% of the variance . The hidden space size of the model is 256, and the discount parameter for the loss function is 0.9.

Generally, \DS  has a smaller Root Mean Squared Error (RMSE) compared to persistence at the all lead times greater than one. Regarding performance with the Global Climate Model (GCM), the median \DS forecast performs better up to month two, but then it is consistently worse than the GCM for the remaining of the forecast. It remains to be understood how this issue can be addressed in future versions of \DS.

\begin{figure*}
 \centering
 \includegraphics[width=39pc]{ 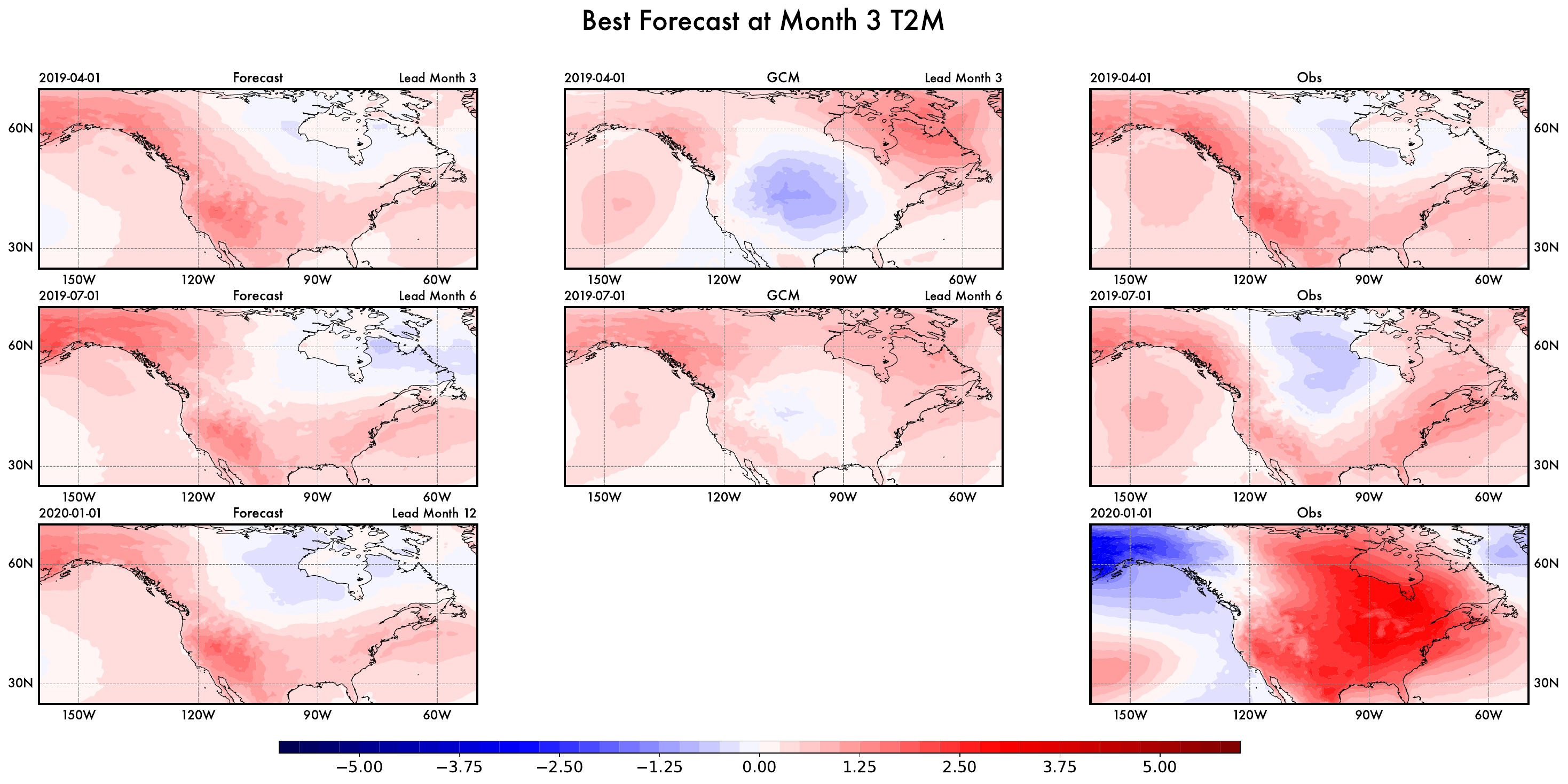}
 \caption{Comparison of anomaly temperature at 2m forecasts from the best-performing \DS configuration and the best score in terms of minimum RMSE among the initial dates at Month three, against observations and the operational GCM. The top set of panels (lead month 3) displays, from left to right, the \DS forecast, the GCM forecast, and the corresponding observations. The middle set (lead month 6) follows the same layout. The bottom set (lead month 9) presents the \DS forecast alongside the observed T2M, omitting the GCM forecast for this lead time that is not available. The color scale (ranging from -5 to 5 ) denotes T2M anomaly values in celsius. } 
\label{Fig:BestNA-T2M}
\end{figure*}

\begin{figure*}
 \centering
 \includegraphics[width=39pc]{ 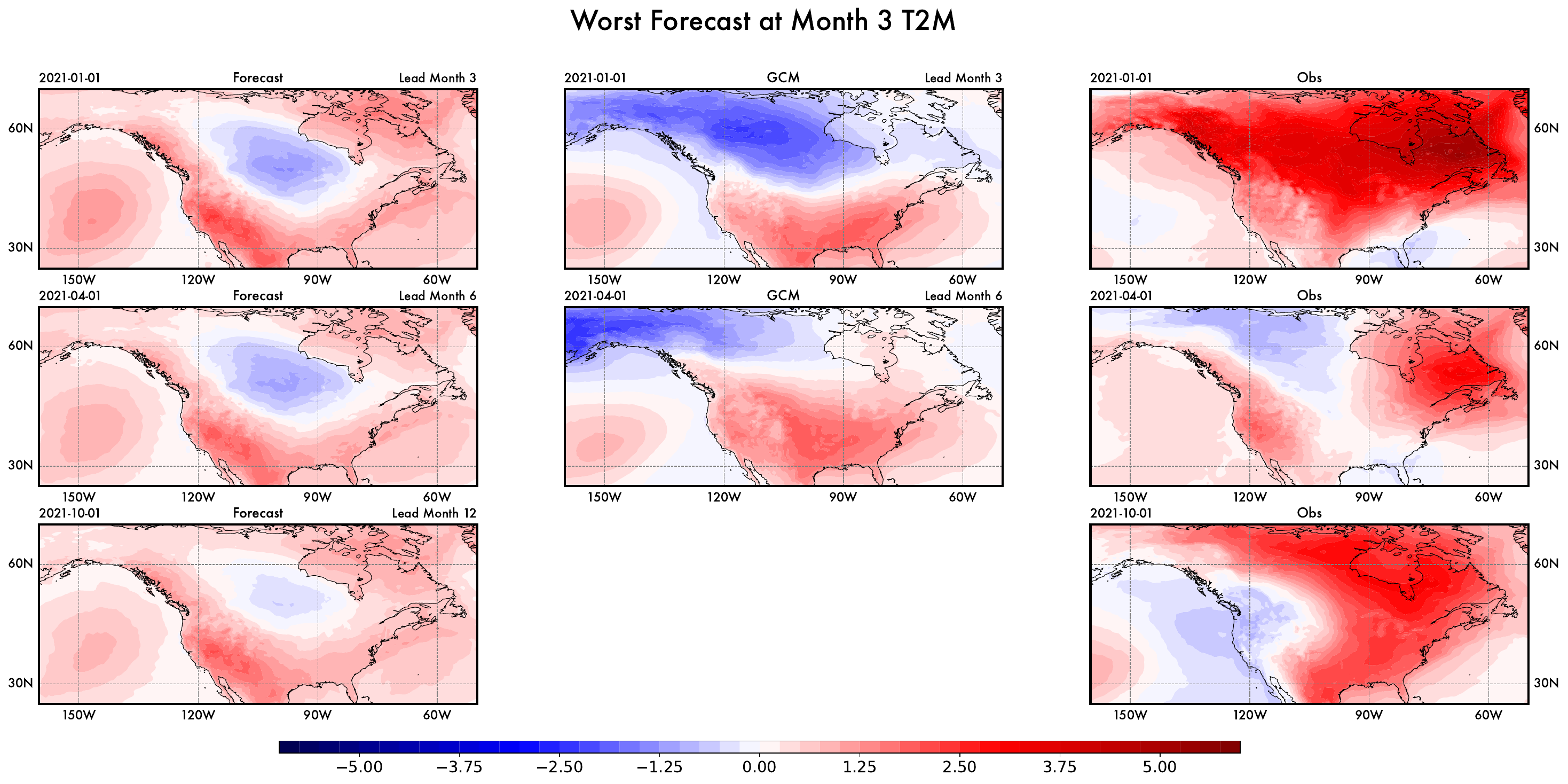}
 \caption{As in Fig. \ref{Fig:BestNA-T2M} but for the worst case.} 
\label{Fig:WorstNA-T2M}
\end{figure*}

The best forecast at month six is shown in Fig.\ref{Fig:BestNA-T2M}.  \DS is capturing the  warm anomalies in the southwest of the continental US and also the colder anomaly over north-east Canada. The evolution from month three to month six is also better described by \DS than the GCM, even if both models are showing a slower evolution than the observations. The worst forecast at month six (Fig.\ref{Fig:WorstNA-T2M}) shows instead a spectacular failure at month three to describe the large warm anomaly over all of North America, failure that is shared with GCM. Clearly the period from October 2010 to January 2011 is a difficult period for the models to forecast.

\begin{figure*}
  \centering
  \includegraphics[width=\textwidth]{ 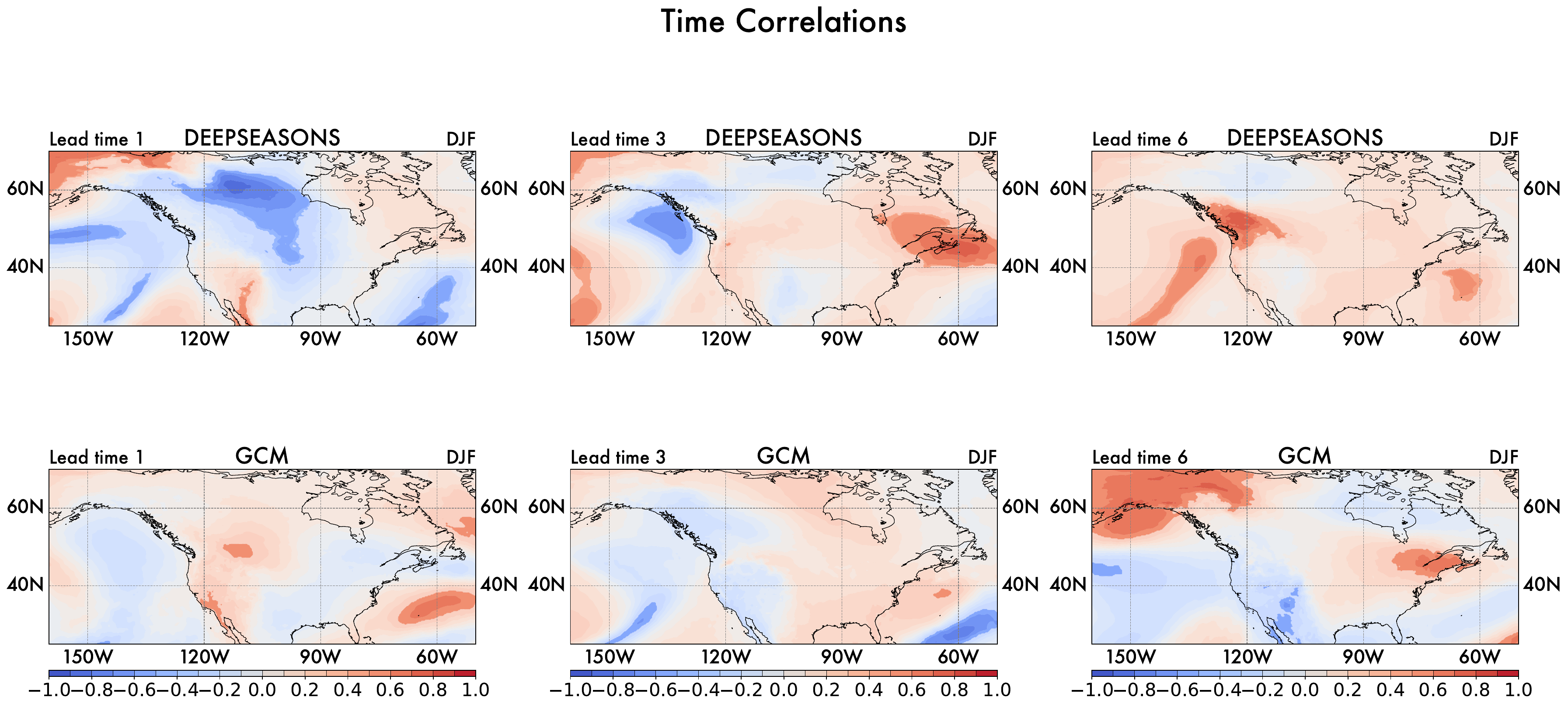}
  \caption{As in Fig.\ref{Fig:SST-Time-Corr} but for North America DJF 2m Temperature. }
  \label{Fig:T2M-NA-DJF-Time-Corr}
\end{figure*}

\begin{figure*}
  \centering
  \includegraphics[width=\textwidth]{ 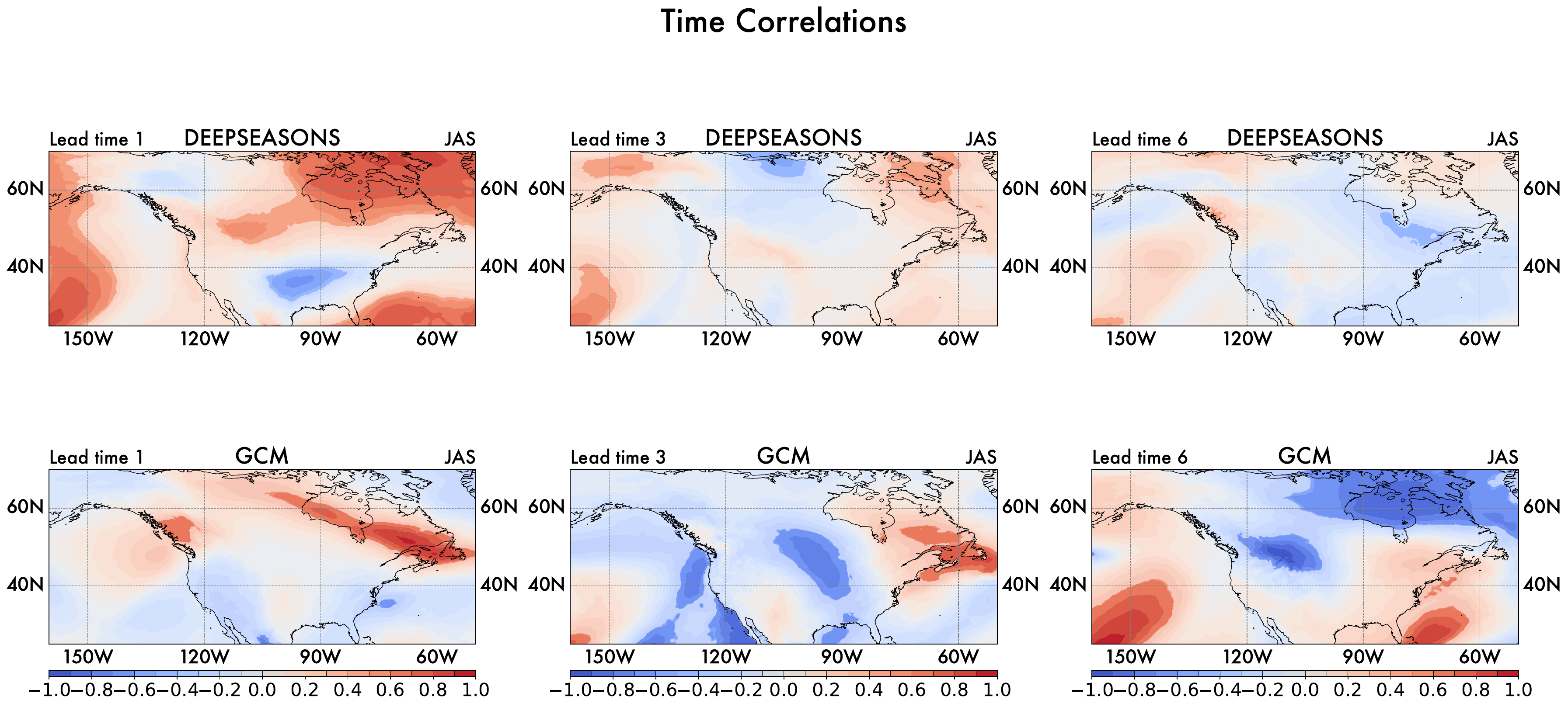}
  \caption{As in Fig.\ref{Fig:SST-Time-Corr} but for North America JAS 2m Temperature. }
  \label{Fig:T2M-NA-JAS-Time-Corr}
\end{figure*}

The potential skill of the model can also be seen in this case, using temporal correlation at a fixed lead time over the ensemble of the forecasts. Figure \ref{Fig:T2M-NA-DJF-Time-Corr} and \ref{Fig:T2M-NA-JAS-Time-Corr} show the results for the North America case. These patterns are difficult to interpret, especially considering that significant negative correlation can be detected at a 10\% significance level. Looking at the general pattern, we can see that \DS, particularly in winter, generally maintains a positive correlation, although the areas of significant value are rather limited. Similar difficulties are observed in traditional GCMs, where we do not have very extensive areas of significant value. However, it is important to note that we are working with only 27 forecasts, and the small number of forecasts may actually bias the evaluation correlation slightly. Some improvements can be seen in the Summer, where \DS have large significant positive correlation at the one month lead time  and generally have good even if not significant correlation at time lead time three. 

\section{Forecasting seasonal (three-months) time averages}

Data-driven methods offer a flexibility that traditional differential equation-based approaches lack, as they do not rely on explicitly solving equations and instead capture relationships directly from the data. While we have previously worked with monthly means, this approach can easily be extended to other time-averaging windows. For instance, instead of focusing on monthly means, we can apply a three-month (seasonal) mean and build a prediction system that targets the average of the upcoming three-month period. Using the data set of monthly means as a base, we can apply a window and generate time series of averages over different periods. We will demonstrate this with a three-month rolling window. In principle, these averages are centered on the window at the center, but for a more operational setting, we adopt a backward-looking approach. This means we assign the mean of the preceding three months to the final month of the window, ensuring that no future information is used in computing the average. Concretely, we implement a rolling three-month average on each dataset and record the resulting average as the value for the last month of the three-month period. Once this preprocessing is done, the data can be handled in the same way as with monthly means. From a mathematical standpoint, the method remains unchanged, allowing us to retain the same neural architecture previously employed.

After a series of heuristic tests on hyper-parameters, such as input variable selection, hidden dimension size, number of Transformer layers, and number of heads, similar to what was performed on the previous monthly mean case, we identified an optimal model configuration. This model uses a 256-dimensional hidden space, one Transformer layer, an 18-time-step input sequence (i.e., 18 three-months means). The input features were selected to represent 90\% of the data variance.

\begin{figure*}[h]
\centering
\includegraphics[width=\textwidth]{ 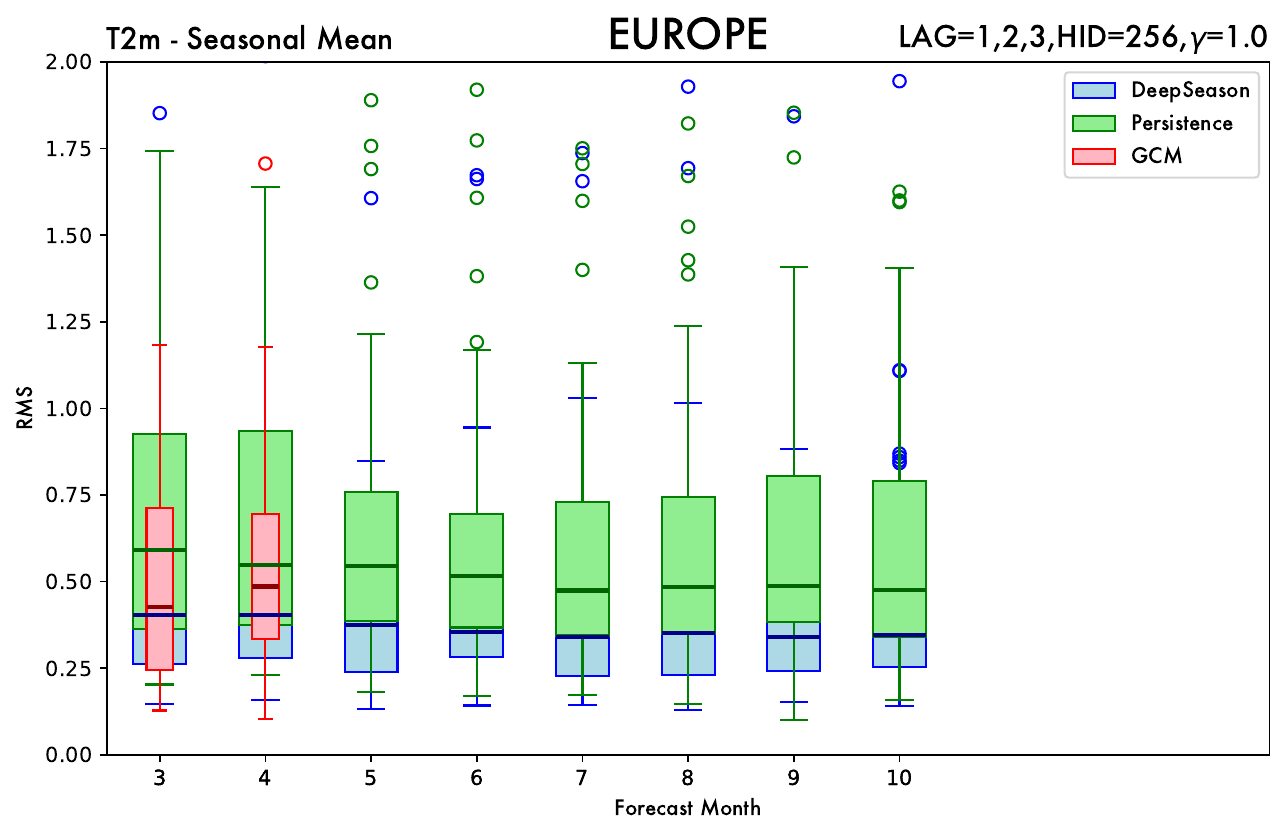}
\caption{RMSE box plot for seasonally averaged forecasts of T2M over the European region.  The plot shows the forecast starting from month Three, as the forecasts are for the average of the preceding three months.
Lightblue indicates the model forecast, lightgreen the persistence forecast, and lightpink the GCM-based forecast.  
Median values appear in dark blue, dark green, and dark red, respectively.
The input fields used are (SST, T2M), and the LAGS parameters are (1, 2, 3, 4).  
In this case, we are keeping enough EOF to represent 90\% in both input fields. We are using here a  hidden space dimension of  256, one transformer layer and a discount parameter of one.}
\label{Fig:Europe-SEAS-Boxplot}
\end{figure*}

After training and validation, forecasts were performed on the same test period of the previous monthly mean case, resulting in 61 forecasts cases. Fig.\ref{Fig:Europe-SEAS-Boxplot} presents a box plot summarizing our results, which show robust predictability of three-month averages using \DS. The figure depicts the root mean square error (RMSE), demonstrating that our method outperforms both a persistence-based baseline and a GCM-based forecast across all points. For consistency, both persistence and GCM forecasts were similarly smoothed (i.e., averaged in three-month windows) to allow a fair comparison.

\begin{figure*}
 \centering
 \includegraphics[width=39pc]{ 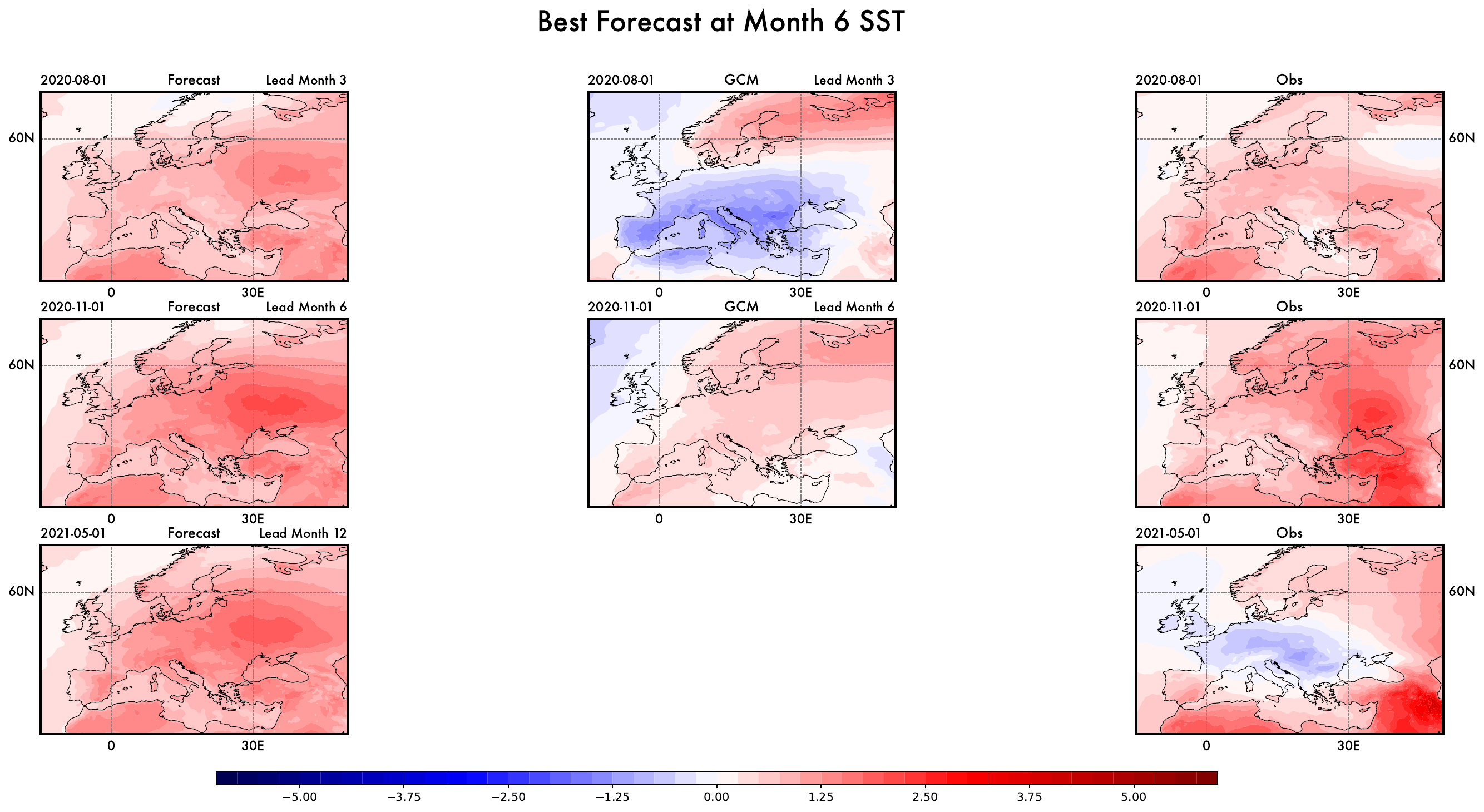}
 \caption{Comparison of seasonally averaged anomaly temperature at 2m forecasts from the best-performing \DS configuration (26 T2M EOF, 14 T850 EOF representing 90\% of the variance, temporal lags of 1-3,  hidden dimension of 256 and  one transformer layer) and the best score in terms of minimum RMSE at Month Three among the initial dates, compared against observations and the operational GCM. The top set of panels (lead month 3) displays, from left to right, the \DS forecast, the GCM forecast, and the corresponding observations. The middle set (lead month 6) follows the same layout. The bottom set (lead month 9) presents the \DS forecast alongside the observed T2M, omitting the GCM forecast for this lead time that is not available. The color scale (ranging from -5 to 5 ) denotes T2M anomaly values in C. } 
\label{Fig:SEASBestT2M}
\end{figure*}

\begin{figure*}
 \centering
 \includegraphics[width=39pc]{ 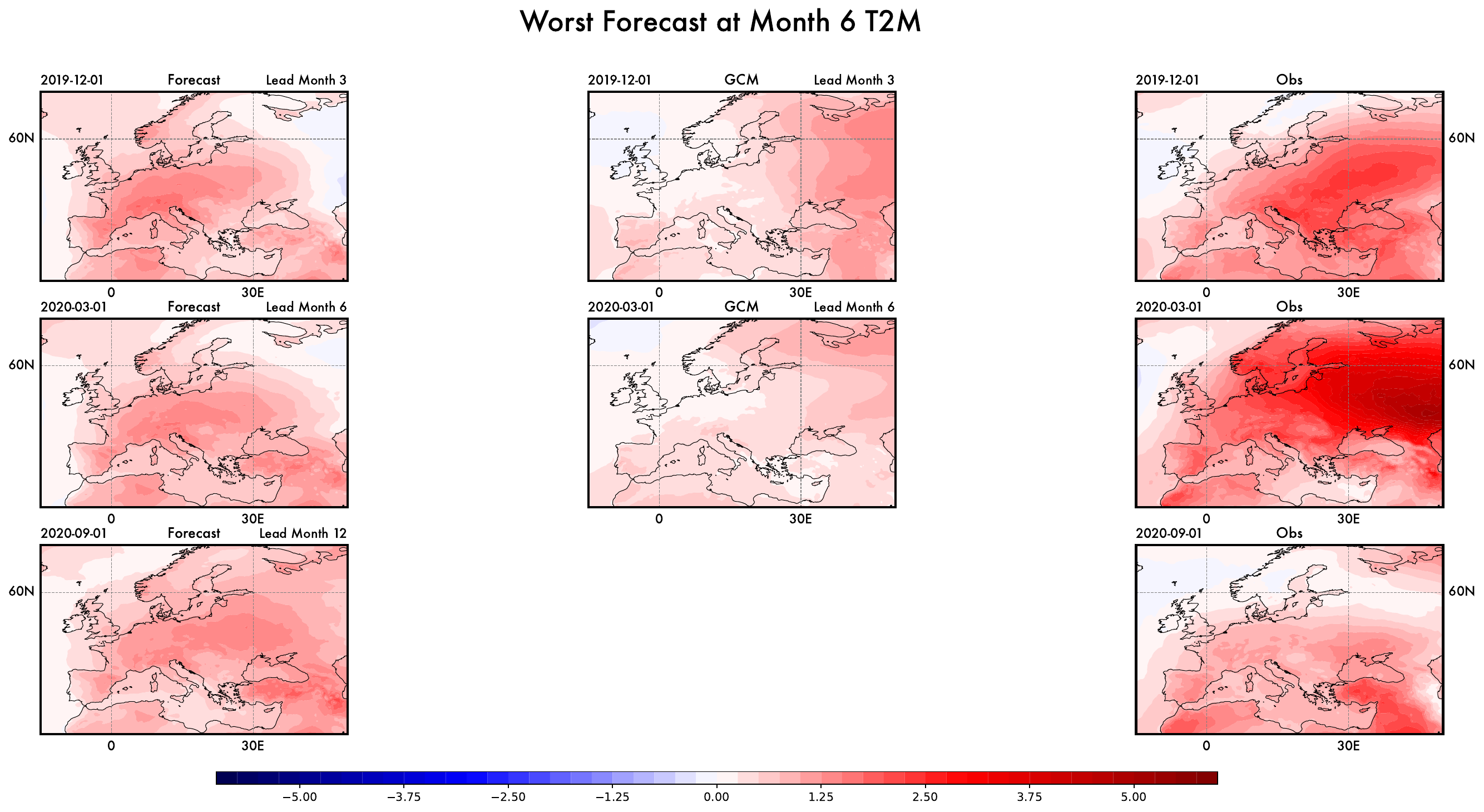}
 \caption{Comparison of seasonally averaged anomaly temperature at 2m forecasts from the worst-performing \DS configuration (26 T2M EOF, 14 T850 EOF representing 90\% of the variance, temporal lags of 1-3,  hidden dimension of 256 and  one transformer layer) and the best score in terms of maximum RMSE at Month Three among the initial dates, compared against observations and the operational GCM. The top set of panels (lead month 3) displays, from left to right, the \DS forecast, the GCM forecast, and the corresponding observations. The middle set (lead month 6) follows the same layout. The bottom set (lead month 9) presents the \DS forecast alongside the observed T2M, omitting the GCM forecast for this lead time that is not available. The color scale (ranging from -5 to 5 ) denotes T2M anomaly values in C. } 
\label{Fig:SEASWorstT2M}
\end{figure*}

The best forecast for the seasonally averaged case is shown in Fig.\ref{Fig:SEASBestT2M}. In this case, \DS succeeds in capturing the positive anomalies over eastern Europe and Russia up to month six realizing a very good performance at his range, but it fails to capture the subsequent development of a cold anomaly over central Europe.   The worst forecast in terms of RMSE (Fig.\ref{Fig:SEASWorstT2M}) is unable to capture the intensification of the anomaly over European Russia. In general \DS exhibits an underestimation of the variability and it does show a tendency to prefer slow evolution.

\begin{figure*}
  \centering
  \includegraphics[width=\textwidth]{ 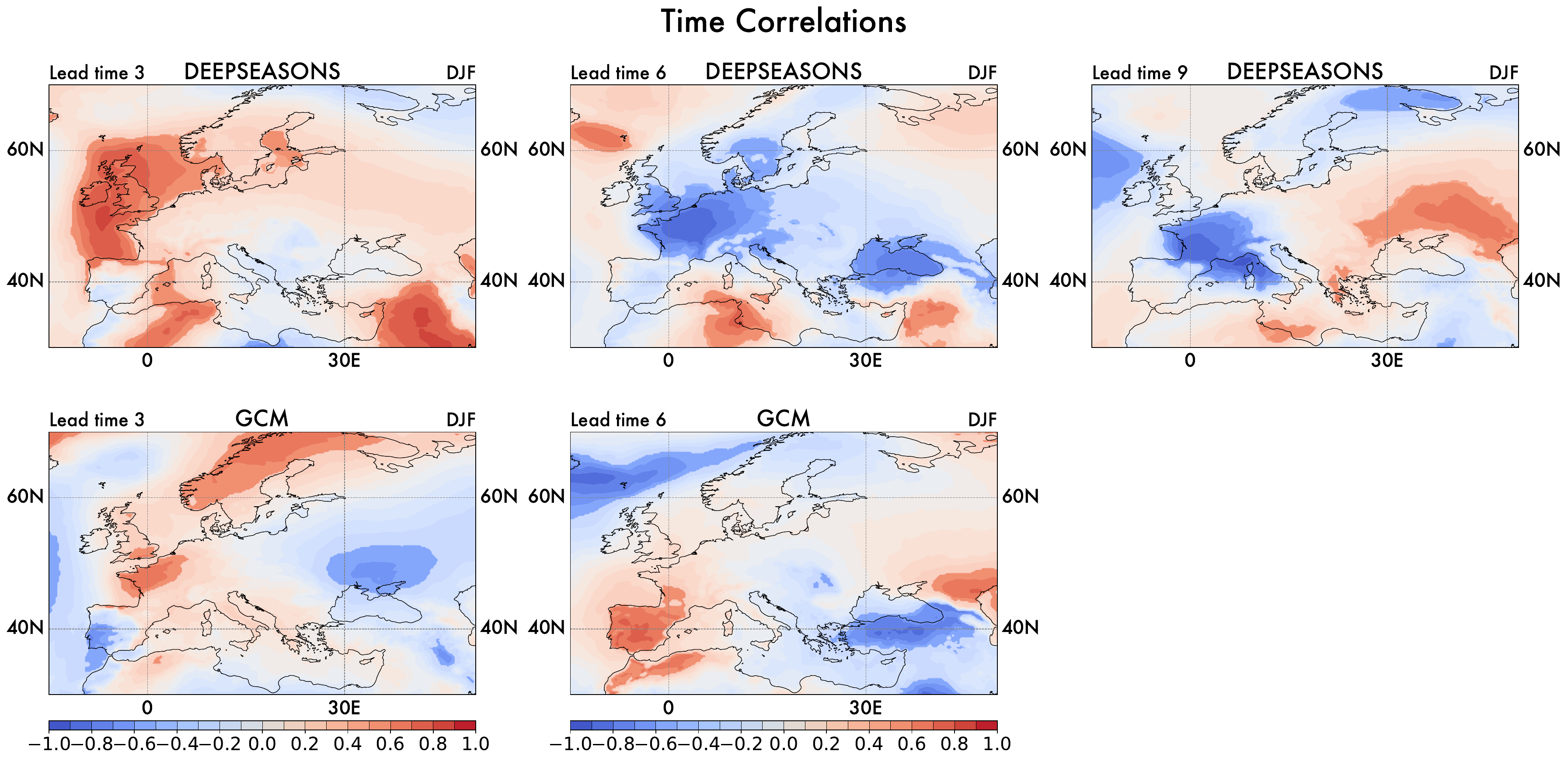}
  \caption{As in Fig.\ref{Fig:SST-Time-Corr} but for European Seasonally (three-months) averages 2m Temperature.}
  \label{Fig:T2M-EU-SEAS-DJF-Time-Corr}
\end{figure*}

The time correlations observed in Winter, as illustrated in Figure \ref{Fig:T2M-EU-SEAS-DJF-Time-Corr}, reveal notably larger areas of significant correlation when compared to the monthly means case. This indicates that the temporal relationships are more pronounced and extend over broader regions during this season. Additionally, these correlations appear to exhibit a more organized structure in comparison to those observed in the General Circulation Model (GCM) case, suggesting an enhanced coherence in the temporal patterns.

\begin{figure*}
  \centering
  \includegraphics[width=\textwidth]{ 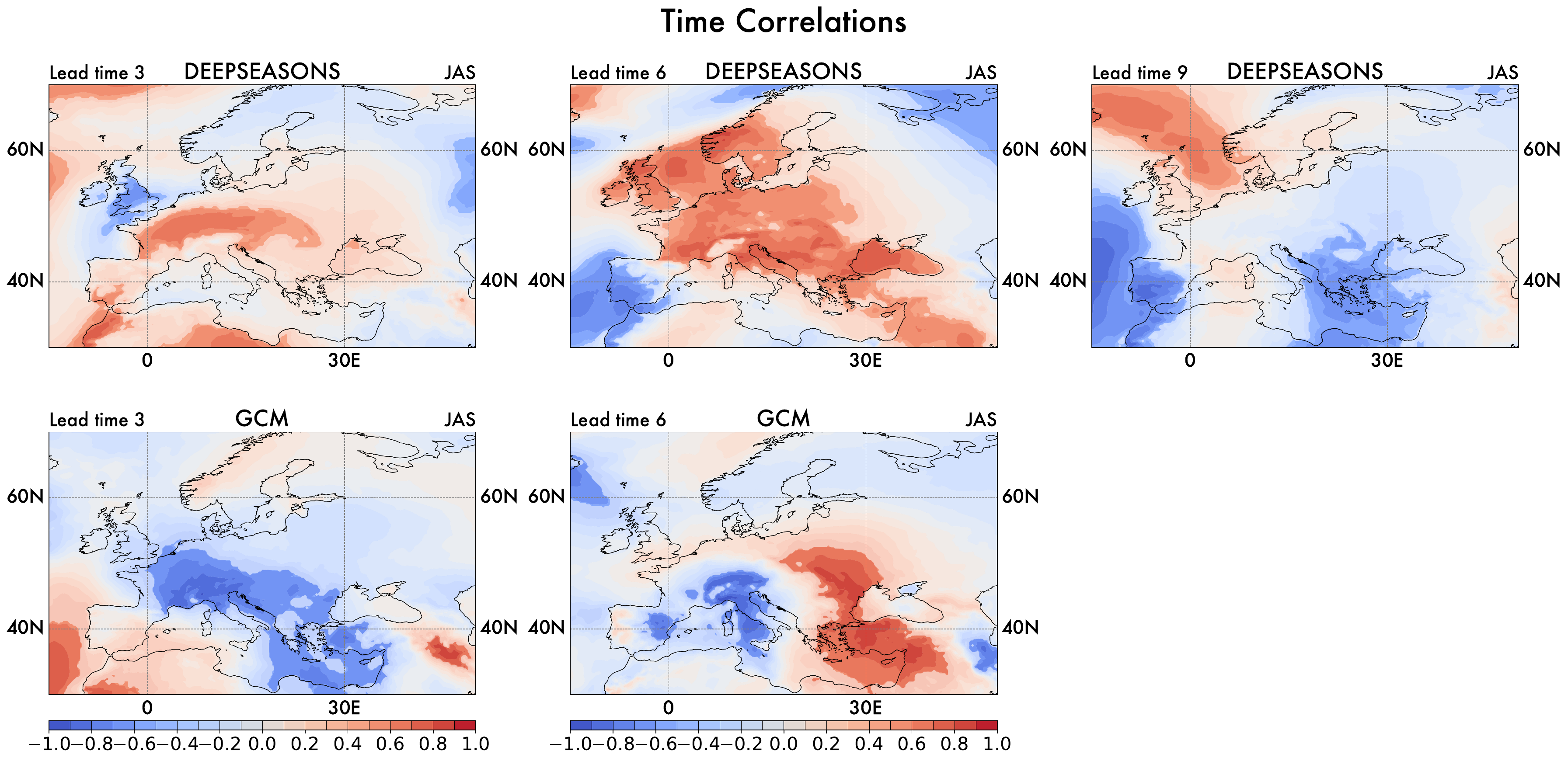}
  \caption{As in Fig.\ref{Fig:T2M-EU-SEAS-DJF-Time-Corr} but for Summer }
  \label{Fig:T2M-EU-SEAS-JAS-Time-Corr}
\end{figure*}

In the Summer season, shown in Figure \ref{Fig:T2M-EU-SEAS-JAS-Time-Corr}, presents a different pattern. Specifically, at the six-month lead time, there are considerably larger areas of positive correlations over time when compared to the GCM. This suggests that the \DS has the ability to capture a better temporal consistency with observations  during this period. Examining other lead seasons, the time correlations remain largely comparable to those observed in the GCM, indicating that while Summer exhibits distinct improvements, the overall temporal correlation performance maintains a level of consistency across different seasons.

\section{Conclusions}

Data-driven methods, by learning directly from historical variability, offer valuable insights into which spatial and temporal scales hold genuine predictive power. Large-scale modes that dominate variance such as ENSO-related patterns in tropical Pacific sea surface temperatures often exhibit greater predictability compared to finer-scale features. By pinpointing areas where models perform optimally, we can strategically allocate computational resources to sectors or modes with the highest intrinsic predictability, rather than distributing efforts uniformly across all regions and timescales. Transforming coordinates to Empirical Orthogonal Functions (EOFs) enhances this process, enabling the selection of dominant, recurrent variability modes that encapsulate the core dynamics of the system. Similarly, focusing on specific time scales helps identify 'islands of predictability,' facilitating more accurate long-term forecasts.

Moreover, aggregating data over extended periods such as monthly or seasonal (e.g., three-month) averages Âreduces noise and accentuates significant patterns. This approach fosters more robust predictions, mitigating the obscuring effects of short-term fluctuations on longer-term trends. Time-averaging also enhances model adaptability, allowing seamless application across diverse temporal scales and operational contexts, whether monthly, seasonal, or annual, depending on specific forecasting needs. Traditional models often require repeated adjustments to correct systematic errors (e.g., model biases in simulating large-scale circulation) and the calculation of very large ensemble to estimate the errors reliably. The correction is applied after the fact, resulting in prediction of anomalies, in this case, departure from the model's climatology.  Machine learning approaches, by contrast, learn directly from historical observations and can be trained specifically on anomalies (departures from observed climatology). This effectively sidesteps many of the corrections demanded by traditional models, as systematic errors are implicitly accounted for in the training data. The result is a more direct path to predicting anomalies, making it easier to capture phenomena that deviate from average conditions.

One of the greatest strengths of purely data-driven approaches is their flexibility. As demonstrated in \DS, these methods can be customized to target not only time-averaged fields but also particular variables of interest, such as regional temperature anomalies, or even precipitation extremes. This tailoring might involve selecting which input variables are fed into the model, how many past time steps to include, or which neural architecture  to use. The end result is a system optimized for the precise forecasting task at hand.

Our research demonstrates that even a relatively straightforward data-driven approach can be effectively designed to rival the predictive capabilities of more complex and comprehensive traditional seasonal forecasting systems. This holds true even when working with the current limitations of available data, such as relying solely on monthly mean analysis data.

While our current model showcases promising forecasting capabilities, there is substantial room for enhancement. Future work will focus on the development of more sophisticated neural network architectures, which will not only improve performance but also enable the extraction of valuable insights from a broader and more diverse array of multivariate input variables and datasets. Despite the simplicity of our current methodology and the constraints of using only monthly means, our findings clearly indicate that state-driven methods possess a genuine potential for accurate and reliable forecasting. Further work will involve exploring how changes in inputs (e.g., different variables or different temporal windows), sequence lengths (the number of past time steps used), or neural network designs (e.g., number of layers, hidden dimension sizes, discount factor and choice of loss function) affect forecast performance. 

This underscores the versatility and potential of data-driven approaches in the realm of seasonal forecasting, paving the way for continued innovation and refinement in predictive modeling techniques.

The potential for further improvements, including other, possibly also model derived, training data set and using other network architecture and design is very large. All calculations for this paper were performed on Apple Mac with M1 or M2 processors from a laptop to a desktop.

\bibliographystyle{plainnat}
\bibliography{SST_AI_no_url}
\end{document}